\documentclass[iop,revtex4,numberedappendix,appendixfloats,twocolappendix]{emulateapj}
\usepackage{times}
\usepackage{appendix}
\usepackage{pdflscape}
\pdfoutput=1

\begin{document}

\newcommand{\oii}{\text{[\ion{O}{2}]}}
\newcommand{\neiii}{\text{[\ion{Ne}{3}]}}
\newcommand{\oiii}{\text{[\ion{O}{3}]}}
\newcommand{\woiii}{\text{$W_\lambda(\oiii)$}}
\newcommand{\nii}{\text{[\ion{N}{2}]}}
\newcommand{\hei}{\text{\ion{He}{1}}}
\newcommand{\heii}{\text{\ion{He}{2}}}
\newcommand{\ha}{\text{H$\alpha$}}
\newcommand{\wha}{\text{$W_\lambda(\ha)$}}
\newcommand{\hb}{\text{H$\beta$}}
\newcommand{\hg}{\text{H$\gamma$}}
\newcommand{\hd}{\text{H$\delta$}}
\newcommand{\he}{\text{H$\epsilon$}}
\newcommand{\hz}{\text{H$\zeta$}}
\newcommand{\hn}{\text{H$\eta$}}
\newcommand{\htheta}{\text{H$\theta$}}
\newcommand{\hiota}{\text{H$\iota$}}
\newcommand{\hi}{\text{\ion{H}{1}}}
\newcommand{\hii}{\text{\ion{H}{2}}}
\newcommand{\hk}{\text{H$\kappa$}}
\newcommand{\caii}{\text{\ion{Ca}{2}}}
\newcommand{\sii}{\text{[\ion{S}{2}]}}
\newcommand{\wlya}{\text{$W_\lambda$({\rm Ly$\alpha$})}}
\newcommand{\wlyaem}{\text{$W_\lambda^{\rm em}$({\rm Ly$\alpha$})}}
\newcommand{\llya}{\text{$L$(Ly$\alpha$)}}
\newcommand{\llyaobs}{\text{$L$(Ly$\alpha$)$_{\rm obs}$}}
\newcommand{\llyaint}{\text{$L$(Ly$\alpha$)$_{\rm int}$}}
\newcommand{\lyafrac}{\text{$f_{\rm esc}^{\rm spec}$(Ly$\alpha$)}}
\newcommand{\lha}{\text{$L$(H$\alpha$)}}
\newcommand{\lhb}{\text{$L$(H$\beta$)}}
\newcommand{\sfrha}{\text{SFR(\ha)}}
\newcommand{\sfrsed}{\text{SFR(SED)}}
\newcommand{\ssfrha}{\text{sSFR(\ha)}}
\newcommand{\ssfrsed}{\text{sSFR(SED)}}
\newcommand{\ebmvneb}{E(B-V)_{\rm neb}}
\newcommand{\ebmvcont}{E(B-V)_{\rm cont}}
\newcommand{\ebmvlos}{E(B-V)_{\rm los}}
\newcommand{\nhi}{N(\text{\ion{H}{1}})}
\newcommand{\lognhi}{\log[\nhi/{\rm cm}^{-2}]}
\newcommand{\lognhitable}{\log\left[\frac{\nhi}{{\rm cm}^{-2}}\right]}
\newcommand{\lya}{\text{Ly$\alpha$}}
\newcommand{\lyb}{\text{Ly$\beta$}}
\newcommand{\lyg}{\text{Ly$\gamma$}}
\newcommand{\comment}[1]{}
\newcommand{\wciii}{\text{$W_\lambda$(\ion{C}{3}])}}
\newcommand{\ciii}{\text{\ion{C}{3}]}}
\newcommand{\interoiii}{\text{\ion{O}{3}]}}
\newcommand{\rsiione}{R(\text{\ion{Si}{2}}\lambda 1260)}
\newcommand{\rsiitwo}{R(\text{\ion{Si}{2}}\lambda 1527)}
\newcommand{\siii}{\text{\ion{Si}{2}}}
\newcommand{\cii}{\text{\ion{C}{2}}}
\newcommand{\civ}{\text{\ion{C}{4}}}
\newcommand{\rcii}{R(\text{\ion{C}{2}}\lambda 1334)}
\newcommand{\ralii}{R(\ion{Al}{2}\lambda 1670)}
\newcommand{\qh}{Q(\text{H$^0$})}
\newcommand{\rs}{{\cal R}_{\rm s}}
\newcommand{\fcovhi}{f_{\rm cov}(\hi)}
\newcommand{\fcovmetal}{f_{\rm cov}({\rm metal})}
\newcommand{\fesclya}{f_{\rm esc}^{\rm spec}(\lya)}
\newcommand{\logxi}{\log[\xi_{\rm ion}/{\rm s^{-1}/erg\,s^{-1}\,Hz^{-1}}]}

\shorttitle{Impact of SFR Surface Density on Electron
Density and Ionization Parameter at High Redshift}
\shortauthors{Reddy et~al.}

\title{The Impact of Star-Formation-Rate Surface Density on the
  Electron Density and Ionization Parameter of High-Redshift
  Galaxies\altaffilmark{*}}

\altaffiltext{*}{Based on data obtained at the W.M. Keck Observatory,
  which is operated as a scientific partnership among the California
  Institute of Technology, the University of California, and NASA, and
  was made possible by the generous financial support of the W.M. Keck
  Foundation.}

\author{\sc Naveen A. Reddy\altaffilmark{1},
Ryan L. Sanders\altaffilmark{2,3},
Alice E. Shapley\altaffilmark{4},
Michael W. Topping\altaffilmark{5},
Mariska Kriek\altaffilmark{6},
Alison L. Coil\altaffilmark{7},
Bahram Mobasher\altaffilmark{1},
Brian Siana\altaffilmark{1},
Saeed Rezaee\altaffilmark{1}}

\altaffiltext{1}{Department of Physics and Astronomy, University of California, 
Riverside, 900 University Avenue, Riverside, CA 92521, USA; naveenr@ucr.edu}
\altaffiltext{2}{Department of Physics, University of California, Davis, One Shields Ave, Davis, CA 95616, USA}
\altaffiltext{3}{NASA Hubble Fellow}
\altaffiltext{4}{Department of Physics \& Astronomy, University of California,
Los Angeles, 430 Portola Plaza, Los Angeles, CA 90095, USA}
\altaffiltext{5}{Steward Observatory, University of Arizona, 933 North 
Cherry Avenue, Tucson, AZ 85721, USA}
\altaffiltext{6}{Leiden Observatory, Leiden University, PO Box 9513, NL-2300 RA Leiden, The Netherlands}
\altaffiltext{7}{Center for Astrophysics and Space Sciences, University of
California, San Diego, 9500 Gilman Drive, La Jolla, CA 92093-0424, USA}

\begin{abstract}

We use the large spectroscopic dataset of the MOSFIRE Deep Evolution
Field (MOSDEF) survey to investigate some of the key factors
responsible for the elevated ionization parameters ($U$) inferred for
high-redshift galaxies, focusing in particular on the role of
star-formation-rate surface density ($\Sigma_{\rm SFR}$).  Using a
sample of 317 galaxies with spectroscopic redshifts $z_{\rm
  spec}\simeq 1.9-3.7$, we construct composite rest-frame optical
spectra in bins of $\Sigma_{\rm SFR}$ and infer electron densities,
$n_e$, using the ratio of the $\oii$\,$\lambda\lambda 3727, 3730$
doublet.  Our analysis suggests a significant ($\simeq 3\sigma$)
correlation between $n_e$ and $\Sigma_{\rm SFR}$.  We further find
significant correlations between $U$ and $\Sigma_{\rm SFR}$ for
composite spectra of a subsample of 113 galaxies, and for a smaller
sample of 25 individual galaxies with inferences of $U$.  The increase
in $n_e$---and possibly also the volume filling factor of dense clumps
in $\hii$ regions---with $\Sigma_{\rm SFR}$ appear to be important
factors in explaining the relationship between $U$ and $\Sigma_{\rm
  SFR}$.  Further, the increase in $n_e$ and SFR with redshift at a
fixed stellar mass can account for most of the redshift evolution of
$U$.  These results suggest that the gas density, which sets $n_e$ and
the overall level of star-formation activity, may play a more
important role than metallicity evolution in explaining the elevated
ionization parameters of high-redshift galaxies.

\end{abstract}

\keywords{stars:abundances --- ISM: abundances --- ISM: HII regions
  --- galaxies: high-redshift --- galaxies: ISM --- galaxies: star
  formation}

\section{\bf INTRODUCTION}
\label{sec:intro}

Spectroscopic surveys of the rest-frame optical nebular emission lines
of redshift $1\la z\la 7$ galaxies have enabled detailed
characterization of the physical state of the interstellar medium
(ISM) and its evolution from the epoch of peak star formation to the
present day.  One important observational finding of these surveys is
the general increase in the ionization parameter (at a fixed stellar
mass) with redshift (e.g., \citealt{brinchmann08, nakajima13,
  steidel14, shirazi14, shapley15, kewley15, bian16, sanders16b,
  kashino17, kojima17, kaasinen18, strom18, topping20a, runco21}),
where the ionization parameter is defined as:
\begin{equation}
U \equiv \frac{n_\gamma}{n_{\rm H}},
\end{equation}
where $n_\gamma$ and $n_{\rm H}$ are the hydrogen-ionizing photon and
hydrogen gas densities, respectively.  For an ionization-bounded
$\hii$ region, $U$ can be written as function of the ionizing photon
rate ($Q$), electron density ($n_e$), and the volume filling factor of
the line-emitting gas ($\epsilon$):
\begin{equation}
U \propto [Q n_e \epsilon^2]^{1/3}
\label{eq:U}
\end{equation}
\citep{charlot01, brinchmann08}.  The redshift
evolution of the ionization parameter at fixed stellar mass is
typically attributed to lower gas-phase oxygen (O) abundances, harder
ionizing spectra (reflective of lower stellar metallicities and/or
younger ages), and/or higher gas (or electron) densities
characteristic of high-redshift galaxies.

In particular, several recent investigations have used
density-sensitive probes---such as the ratios of the $\oii$ or $\sii$
doublet lines---to infer electron densities ($n_{e}$) that are
elevated by up to an order of magnitude at $z\sim 1-2$ relative to
typical star-forming galaxies in the local universe (e.g.,
\citealt{lehnert09, masters14, steidel14, shimakawa15, bian16,
  steidel16, sanders16b, davies21}).  The apparent evolution of $n_e$
may be tied to the higher star-formation rates (SFRs), specific SFRs,
and/or SFR surface densities ($\Sigma_{\rm SFR}$) of high-redshift
galaxies.  For example, \citet{brinchmann08}, \citet{liu08}, \citet{masters16}, and
\citet{bian16} found that nearby galaxies that are offset in the same
direction as high-redshift galaxies from the local star-forming
sequence in the $\oiii/\hb$ versus $\nii/\ha$ BPT plane exhibit
higher $\Sigma_{\rm SFR}$ and/or $n_{e}$.  Further, \citet{kaasinen17}
highlighted the similarity in $n_{e}$ of local and high-redshift
galaxies when matched in SFR.  \citet{shimakawa15} used a small sample
of 14 $\ha$ emitters at $z=2.5$ to directly demonstrate a significant
correlation between $n_e$ and $\Sigma_{\rm SFR}$, one that has been
subsequently confirmed to exist for local analogs of high-redshift
Ly$\alpha$ emitters (``Green Pea'' galaxies) and local Lyman-break
analogs (\citealt{jiang19}; see also \citealt{herrera16}).  Similarly,
based on data from the KMOS-3D survey, \citet{davies21} suggest that
the redshift evolution of $n_e$ may be tied to the increasing density
of molecular clouds.

Thus, aside from an increase in ionizing photon rates, the higher
$\Sigma_{\rm SFR}$ (or molecular gas densities) characteristic of
high-redshift galaxies, accompanied by higher $n_{e}$, may be partly
responsible for the elevated ionization parameters inferred for $z\sim
2$ galaxies (e.g., \citealt{shirazi14, bian16, reddy22}).
Unfortunately, the physical interpretation of correlations between
ionization parameter and other global galaxy properties (e.g.,
$\Sigma_{\rm SFR}$) is complicated by the fact that the ionization
parameter is not directly observable.  Rather, it is usually inferred
using photoionization modeling with simplified (plane-parallel or
spherical) geometries that do not faithfully capture the complicated
structure of real $\hii$ regions (e.g., \citealt{pellegrini11}; see
also discussion in \citealt{sanders16b}).  As a result, the ratio of
emission lines from two different ionization stages of the same
element is often used as proxy for the ionization parameter.
Examining such line ratios can give useful insights into the primary
factors that modulate the ionization parameter in high-redshift
galaxies.

In this paper, we use the MOSFIRE Deep Evolution Field (MOSDEF)
spectroscopic survey data \citep{kriek15} in the CANDELS fields
\citep{grogin11, koekemoer11} in addition to predictions of
photoionization models to investigate a few of the relevant factors
responsible for the elevated ionization parameters inferred for
high-redshift galaxies.  The MOSDEF survey is particularly well-suited to
address this issue since it targets many of the strong rest-frame
optical emission lines that probe the ionization parameter,
gas-phase oxygen abundance, and electron density.  In addition,
rest-frame FUV spectroscopy of a subsample of MOSDEF galaxies
(MOSDEF-LRIS; \citealt{topping20a, reddy22}) enables direct
constraints on the hardness of the ionizing spectra of the same
galaxies.  Finally, the deep {\em HST} imaging that exists in the
CANDELS fields enables measurements of $\Sigma_{\rm SFR}$.  All these
elements together form an ideal dataset with which to study the
evolving relationships between stellar metallicities, ages, SFRs,
stellar masses, gas-phase abundances, and ionization parameters
\citep{shapley15, sanders15, sanders16b, sanders18, topping20a,
  runco21}.  

Here, we extend these previous efforts by focusing on how electron
densities and ionization parameters correlate with $\Sigma_{\rm SFR}$,
and investigating the relative importance of electron or gas density
in explaining the variation in ionization parameters inferred for
high-redshift galaxies.  Section~\ref{sec:sample} summarizes the
MOSDEF survey and the samples analyzed in this work.
Sections~\ref{sec:electrondensity} and \ref{sec:ionizationparameter}
present our findings regarding correlations between electron density
and $\Sigma_{\rm SFR}$, and between ionization parameter and
$\Sigma_{\rm SFR}$, respectively.  The implications of our results for
the variation in $U$ among galaxies in our sample, and the redshift
evolution of $U$, are discussed in Section~\ref{sec:discussion}.  The
conclusions are presented in Section~\ref{sec:conclusions}.  A
\citet{chabrier03} initial mass function (IMF) is considered
throughout the paper.  Wavelengths are reported in the vacuum frame.
We adopt a cosmology with $H_{0}=70$\,km\,s$^{-1}$\,Mpc$^{-1}$,
$\Omega_{\Lambda}=0.7$, and $\Omega_{\rm m}=0.3$.

\section{\bf SAMPLE}
\label{sec:sample}

The galaxies analyzed here were drawn from the MOSDEF survey
\citep{kriek15}.  This survey targeted $\approx 1500$ {\em
  H}-band-selected galaxies and AGNs at redshifts $1.4\la z\la 3.8$ in
the CANDELS fields \citep{grogin11, koekemoer11} with moderate
resolution ($R\sim 3000-3600$) rest-frame optical spectroscopy using
the MOSFIRE spectrometer \citep{mclean12} on the Keck telescope.
Details of the survey, spectroscopic data reduction, and line flux
measurements are provided in \citet{kriek15} and \citet{reddy15}.

The analysis of electron densities (Section~\ref{sec:electrondensity})
is based on the subset of MOSDEF galaxies with secure spectroscopic
redshifts $z_{\rm spec}\ge 1.9$\footnote{This lower limit on the
  spectroscopic redshift ensures that the $\oii$ doublet is
  sufficiently resolved in the observed frame to reliably determine
  the ratio of the doublet lines.}; no evidence of AGN based on the
criteria of \citet{coil15}, \citet{azadi17, azadi18}, and
\citet{leung19}; spectral coverage and no significant sky line
contamination of $\oii$\,$\lambda\lambda 3727,3730$\footnote{While
  $\sii$\,$\lambda\lambda 6718, 6733$ can also be used to infer $n_e$,
  the weakness of this doublet relative to $\oii$ results in less
  stringent constraints on $n_e$, and therefore we chose to focus on
  $\oii$.}; and reliable half-light radii, $R_{\rm eff}$, and their
measurement uncertainties based on the size catalogs of
\citet{vanderwel14}.  These criteria result in a ``density'' sample
consisting of 317 galaxies which span the full range of SFR and
$M_\ast$ as the parent MOSDEF sample.

\begin{figure}
  \epsscale{1.2}
  \plotone{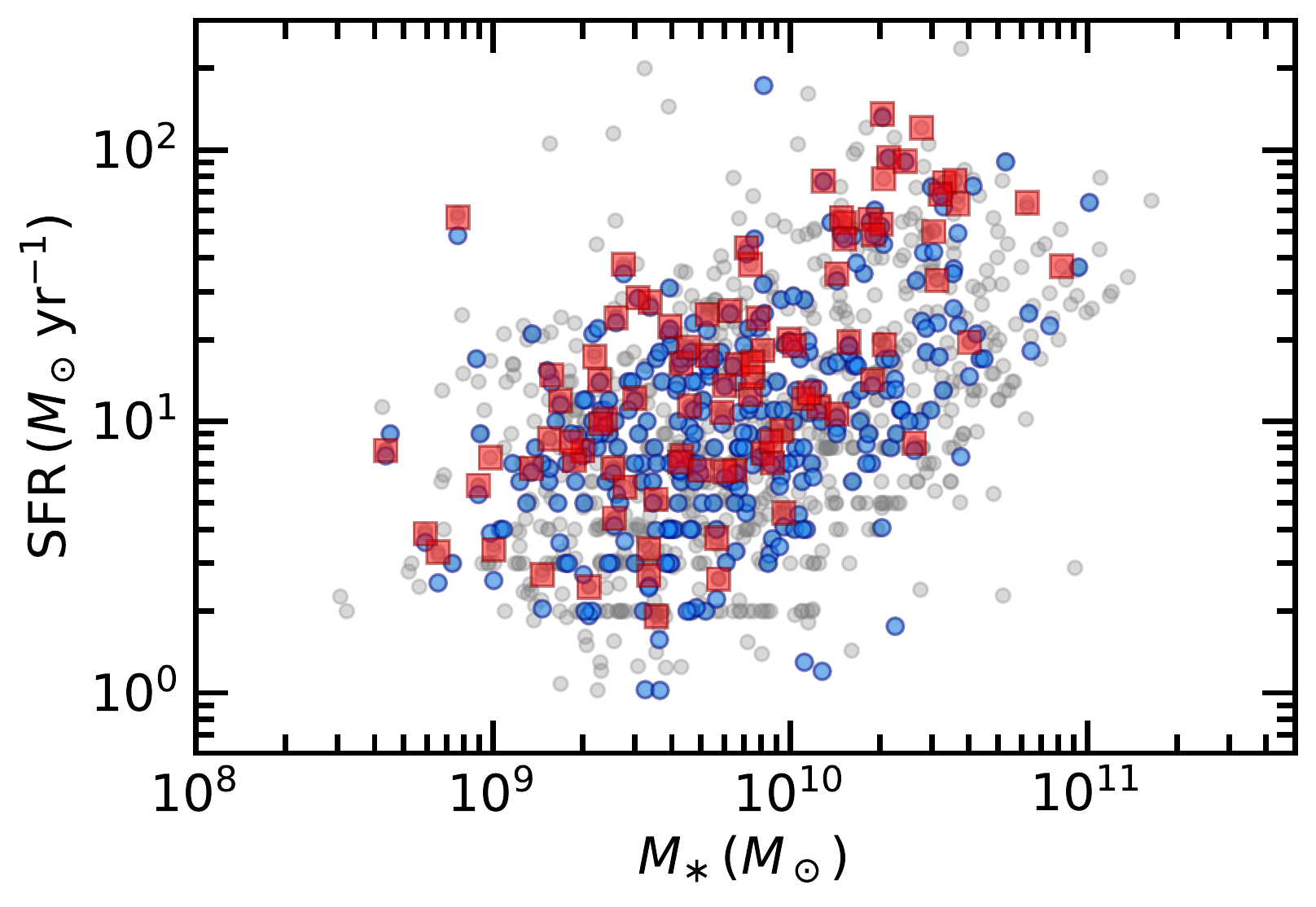}
    \caption{Distributions of SFR and $M_\ast$ for the density and
      ionization parameter samples (blue and red symbols,
      respectively) relative to those of the parent MOSDEF sample
      (grey symbols).  The SFRs shown here are mostly based on the
      $\ha$ luminosities corrected for dust based on the Balmer
      decrement (e.g., \citealt{reddy15}).  If such measurements are
      not available (e.g., $\ha$ is not covered and/or $\ha$ or $\hb$
      are not significantly detected), then the SFRs are based on
      modeling the broadband photometry (Section~\ref{sec:sample}).}
\label{fig:sfrmstar}
\end{figure}

The analysis of the ionization parameter
(Section~\ref{sec:ionizationparameter}) is based on the subset of
MOSDEF galaxies with secure spectroscopic redshifts $1.59< z_{\rm
  spec}< 2.56$; no evidence of AGN based on the criteria of
\citet{coil15}, \citet{azadi17, azadi18}, and \citet{leung19};
spectral coverage and no significant sky line contamination of $\oii$,
$\oiii$, $\hb$, and $\ha$\footnote{The requirement for coverage of
  $\ha$ and $\hb$ ensures that the ratio of $\oiii$ to $\oii$ can be
  robustly corrected for dust attenuation.}; and reliable half-light
radii, $R_{\rm eff}$, and their measurement uncertainties based on the
size catalogs of \citet{vanderwel14}.  These criteria result in an
``ionization parameter'' sample consisting of 113 galaxies which span
the full range of SFR and $M_\ast$ as the parent MOSDEF sample.  For
reference, the distributions of SFR and $M_\ast$ for galaxies in the
density and ionization parameter samples relative to those of the
parent MOSDEF sample are shown in Figure~\ref{fig:sfrmstar}.

The $R_{\rm eff}$ for galaxies in the two aforementioned samples,
along with SFRs calculated from fitting the broadband spectral energy
distributions (SEDs) of the galaxies, are used to compute $\Sigma_{\rm
  SFR}$ as described in \citet{reddy22}.  The SED-inferred SFRs assume
the same Binary Population and Spectral Synthesis (BPASS;
\citealt{eldridge17}) models discussed in \citet{reddy22}, where we
adopted an SMC attenuation curve for the reddening of the stellar
continuum.\footnote{Assuming the \citet{calzetti00} curve for all
  galaxies, or the SMC and \citet{calzetti00} curves for low- and
  high-mass galaxies, respectively, e.g., as suggested by
  \citet{shivaei20}, does not alter our conclusions.  We have adopted
  an SMC curve as this choice provides the best agreement between
  $\ha$ and UV SFRs \citep{reddy18b, reddy22}---and reproduces the
  observed IRX-$\beta$ relation at $z\sim 2$ \citep{reddy18a}---for
  subsolar-stellar-metallicity and/or young stellar populations (see
  also \citealt{reddy06a, shivaei15a, theios19}).}

\begin{deluxetable}{lcccc}
\tabletypesize{\footnotesize}
\tablewidth{0pc}
\tablecaption{Density Subsamples}
\tablehead{
\colhead{Subsample} &
\colhead{$N$\tablenotemark{a}} &
\colhead{$\langle \Sigma_{\rm SFR}\rangle$\tablenotemark{b}} &
\colhead{$\langle R\rangle$\tablenotemark{c}} &
\colhead{$\langle n_e \rangle$\tablenotemark{d}}
\\
 & & $(M_\odot\,{\rm yr}^{-1}\,{\rm kpc}^{-2})$ & & $({\rm cm}^{-3})$}
\startdata
   $\Sigma_{\rm SFR}$,Q1 & 79 & $0.107\pm0.005$ & $1.326\pm0.101$ & $88^{+87}_{-70}$ \\
   $\Sigma_{\rm SFR}$,Q2 & 79 & $0.260\pm0.010$ & $1.231\pm0.064$ & $170^{+66}_{-57}$ \\
   $\Sigma_{\rm SFR}$,Q3 & 79 & $0.453\pm0.013$ & $1.247\pm0.068$ & $155^{+67}_{-58}$ \\
   $\Sigma_{\rm SFR}$,Q4 & 80 & $1.991\pm0.358$ & $1.026\pm0.066$ & $425^{+121}_{-98}$ \enddata
\tablenotetext{a}{Number of galaxies in the subsample.}
\tablenotetext{b}{Average star-formation-rate surface density.}
\tablenotetext{c}{Average $\oii$ line ratio.}
\tablenotetext{d}{Average electron density.}
\label{tab:densitysubsamples}
\end{deluxetable}

\begin{figure}
  \epsscale{1.2}
  \plotone{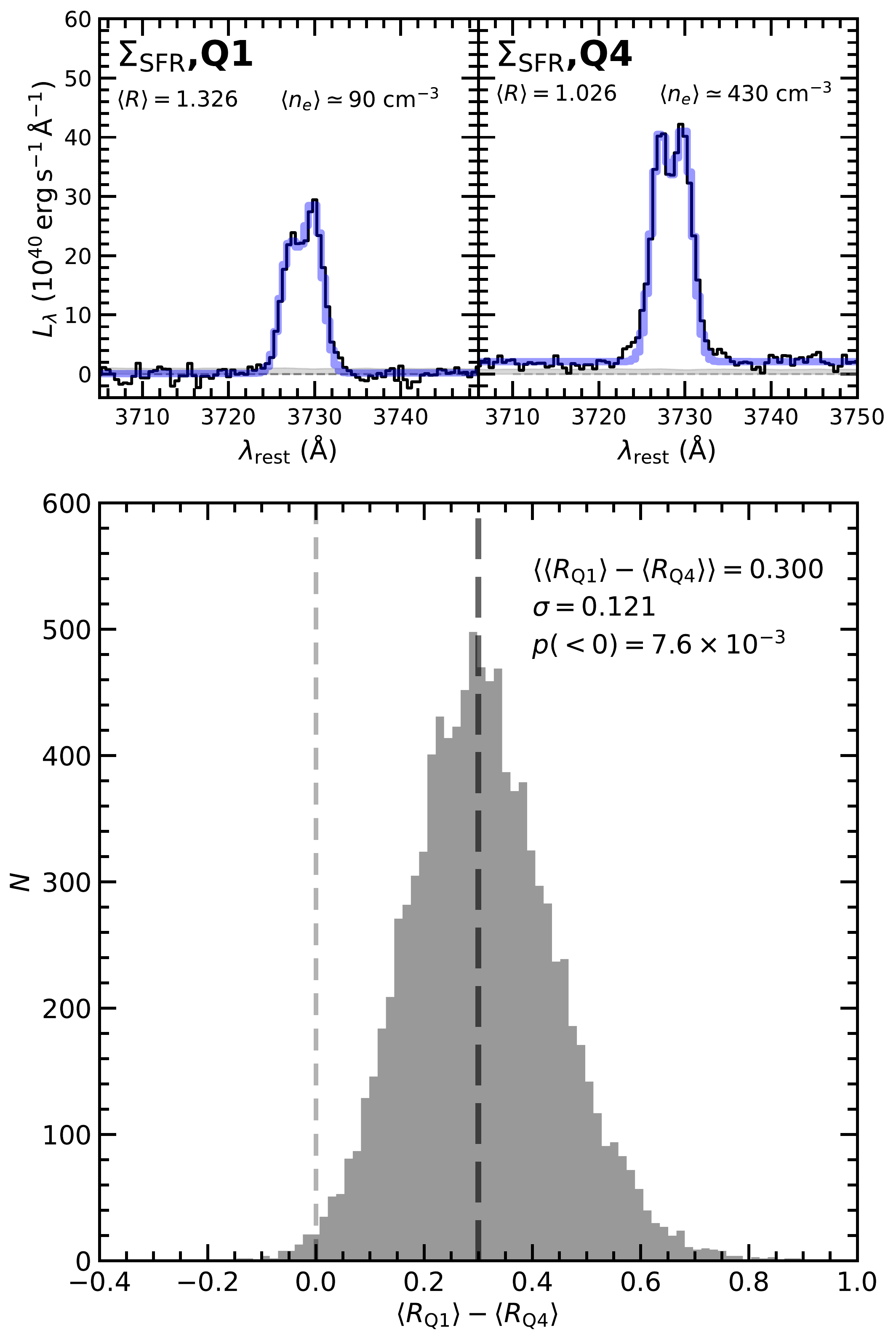}
    \caption{Top: Composite spectra of the $\oii$\,$\lambda\lambda
      3727,3730$ doublet for galaxies in the lowest (left) and highest
      quartiles (right) of $\Sigma_{\rm SFR}$.  The grey shaded
      regions indicate the composite error spectra.  Model fits to the
      $\oii$ doublet are indicated in blue, with the average line
      ratio, $\langle R\rangle = \langle \oii\,\lambda 3730 /
      \oii\,\lambda 3727\rangle$, and inferred density, $\langle
      n_e\rangle$, indicated in each panel. Bottom: Distribution of
      the difference in line ratios measured from 10,000 realizations
      of the composite spectra in the lowest and highest quartiles of
      $\Sigma_{\rm SFR}$.  The long-dashed line indicates the average
      difference of $\langle\langle R_{\rm Q1}\rangle - \langle R_{\rm
        Q4}\rangle\rangle = 0.300$.  Also indicated are the standard
      deviation ($\sigma$) and the probability that the average line
      ratios measured for the two quartiles are equivalent.}
    \label{fig:density1}
\end{figure}

\section{\bf ELECTRON DENSITY}
\label{sec:electrondensity}

Electron densities were inferred from the ratio of $\oii$\,$\lambda
3730$ to $\oii$\,$\lambda 3727$ ($R$) as described in
\citet{sanders16b}.  Robust $n_e$ constraints on individual galaxies
require extremely high $S/N$ measurements of $\oii$, higher than what
is typically available in the MOSDEF spectra.  As a result, we focused
on measuring $R$ from composite spectra to obtain the tightest
constraints on $n_e$.  The density sample was divided into four bins
of $\Sigma_{\rm SFR}$, each containing roughly an equal number of
galaxies.  Composite spectra were constructed for galaxies in each of
these bins using the methodology described in \citet{reddy22}.
Specifically, individual galaxy spectra were averaged together
assuming no weighting in order to avoid biasing the composite to the
more luminous galaxies in the sample.  The strong rest-frame optical
emission lines in the composite spectra for the four bins of
$\Sigma_{\rm SFR}$ are shown in Appendix~\ref{sec:appA}.  $R$ was
measured by fitting simultaneously two Gaussian functions to the two
lines of the $\oii$ doublet assuming an intrinsic line width
equivalent to that measured for $\oiii$\,$\lambda 5008$, and
calculating the ratio of the $\oii$\,$\lambda 3730$ line flux to the
$\oii$\,$\lambda 3727$ line flux.  Uncertainties in $R$ were
determined by perturbing individual science spectra by the
corresponding error spectra, reconstructing the composite spectra from
these individual perturbed spectra many times with replacement, and
remeasuring $R$.

\begin{figure}
  \epsscale{1.2}
  \plotone{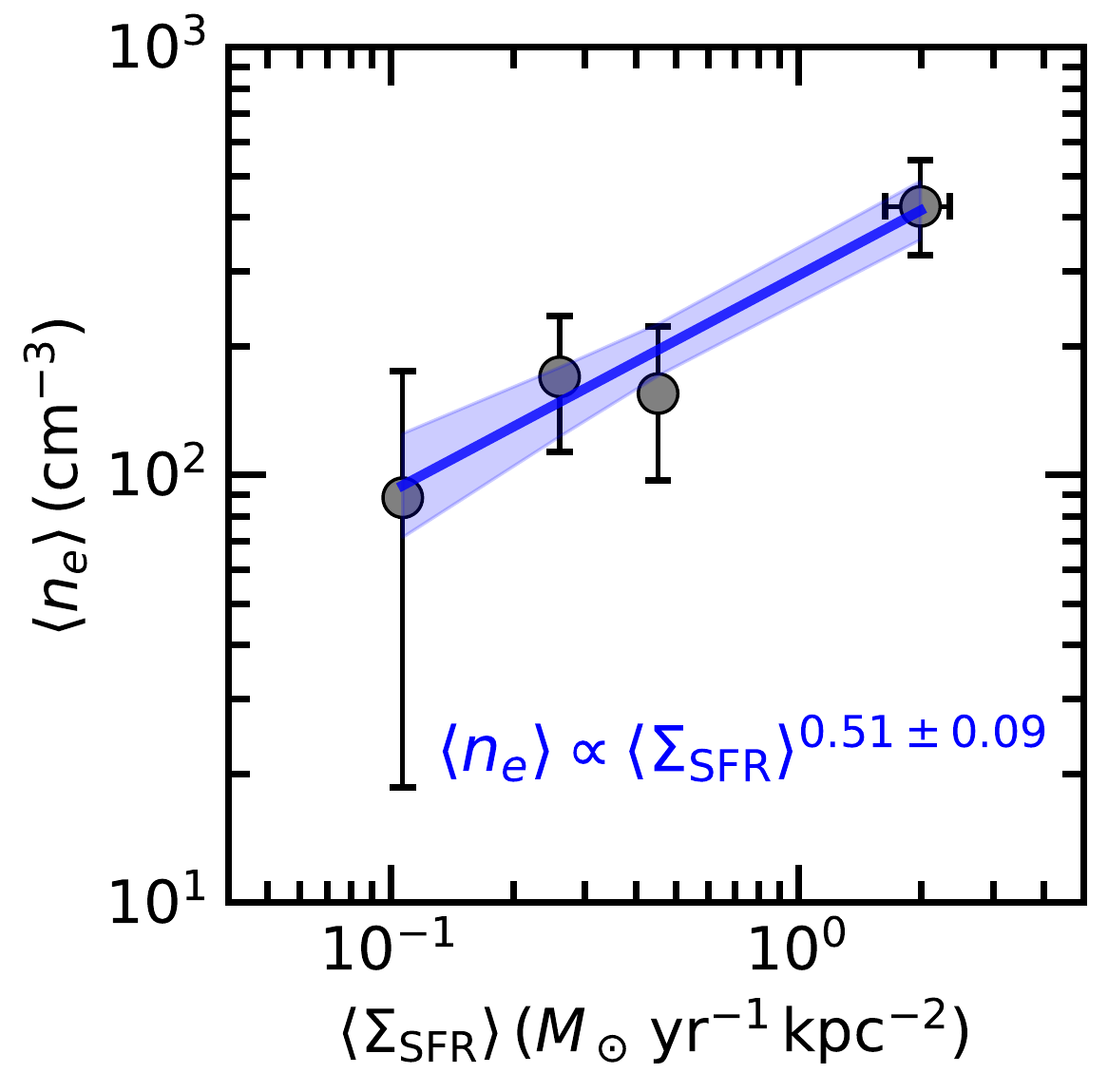}
    \caption{Average electron density versus average $\Sigma_{\rm
        SFR}$ for the four quartiles of $\Sigma_{\rm SFR}$.  The
      best-fit power law and $1\sigma$ confidence interval are
      indicated by the solid blue line and shaded blue region,
      respectively.}
    \label{fig:density2}
\end{figure}

Table~\ref{tab:densitysubsamples} lists the number of galaxies in each
of the four subsamples, along with the average $\Sigma_{\rm SFR}$,
$R$, and the inferred $n_{e}$ for each subsample.  The top panels of
Figure~\ref{fig:density1} illustrate the $\oii$ fits obtained for the
bottom and top quartile bins of $\Sigma_{\rm SFR}$.  The bottom panel
of this figure shows the distribution of the difference in $\langle
R\rangle$ for the bottom and top quartile bins of $\Sigma_{\rm SFR}$
obtained from 10,000 realizations of the data.  A one-sided $t$ test
of this distribution indicates a probability $p\simeq 7.6\times
10^{-3}$ that there is no statistical difference in the $\oii$ doublet
line ratio for the bottom and top quartile bins of $\Sigma_{\rm SFR}$.
Electron densities were calculated from the line ratios using the
prescription given in \citet{sanders16b}, and are shown in
Figure~\ref{fig:density2}.

Our results imply a significant and close to a factor of $5$ increase
in $\langle n_e\rangle$ for galaxies in the top quartile of
$\Sigma_{\rm SFR}$ relative to those in the bottom quartile.  A formal
fit to $\langle n_e\rangle$ versus $\langle\Sigma_{\rm SFR}\rangle$
for the four bins of $\Sigma_{\rm SFR}$ implies
\begin{equation}
\langle n_e\rangle \propto \langle\Sigma_{\rm SFR}\rangle^{0.51\pm 0.09}
\label{eq:nesigmasfr}
\end{equation}
(Figure~\ref{fig:density2}), consistent with the power-law index of
$\approx 0.61$ found by \citet{shimakawa15} for a much smaller sample
of 14 individual $\ha$ emitters at $z=2.5$.  The power-law index found
here is also consistent (within the $68\%$ confidence intervals) of
the index obtained from fitting individual local Lyman break galaxy
analogs and Ly$\alpha$ emitters \citep{jiang19}.  Note that the
scaling relation of Equation~\ref{eq:nesigmasfr} is based on {\em
  composite} measurements of $n_e$.  It is possible that outliers in
the lowest and highest bins are averaged out in the composites in a
way that produces a shallower power-law index than what would have
been obtained from fitting individual measurements of $n_e$.
Consequently, it is possible that the power-law index between $n_e$
and $\Sigma_{\rm SFR}$ may be larger than the value found here,
strengthening our conclusion of a significant correlation between
$n_e$ and $\Sigma_{\rm SFR}$.  High signal-to-noise measurements of
density-sensitive indicators such as $\oii$ for individual galaxies
may be necessary to robustly constrain the power-law index.  At any
rate, the dependence of ionization parameter on $n_e$
(Section~\ref{sec:ionizationparameter}), the positive correlation
between $n_e$ and $\Sigma_{\rm SFR}$, and the generally higher
$\Sigma_{\rm SFR}$ characteristic of high-redshift galaxies, together
may provide an explanation for the elevated ionization parameters
inferred at high redshift, a point to which we return to below.

\section{\bf IONIZATION PARAMETER}
\label{sec:ionizationparameter}

The ratio of $\oiii$ to $\oii$, i.e.,
\begin{equation}
{\rm O32} = \log\left[\frac{\oiii\,\lambda 4960 + \oiii\,\lambda 5008}
{\oii\,\lambda 3727 + \oii\,\lambda 3730}\right]
\end{equation}
is commonly used as a proxy for the ionization parameter, $U$ (e.g.,
\citealt{nakajima14}).  Here, we investigate the correlation between
$U$ and $\Sigma_{\rm SFR}$, first using O32 as a proxy for the former,
and then using estimates of $U$ based on photoionization modeling of
individual galaxies.

\begin{figure}
  \epsscale{1.2}
   \plotone{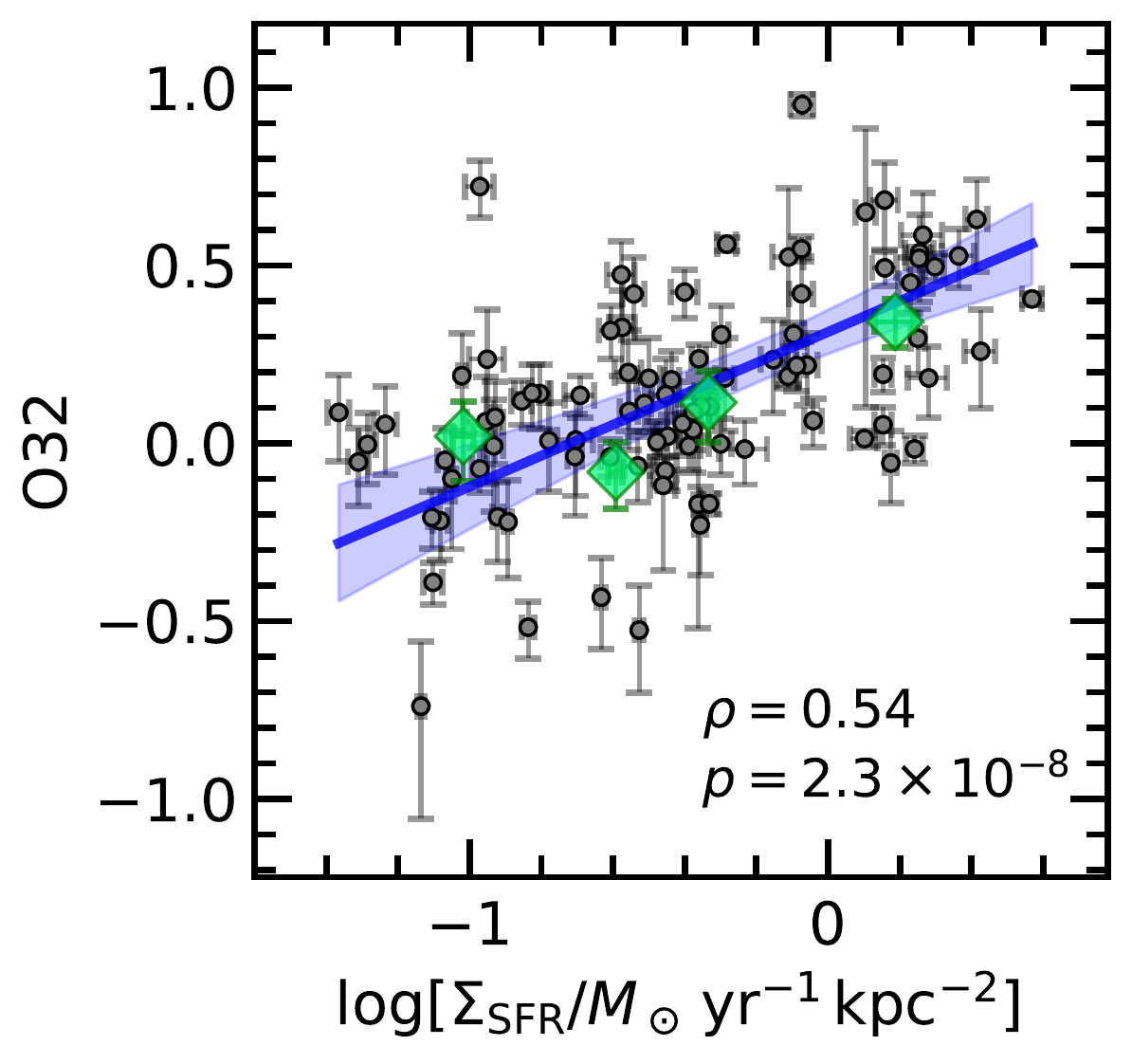}
    \caption{O32 versus $\log[\Sigma_{\rm SFR}/M_\odot\,{\rm
          yr}^{-1}\,{\rm kpc}^{-2}]$ for 93 galaxies in the
      ``ionization parameter'' sample with $S/N\ge 3$ detections of
      $\oii$, $\oiii$, $\hb$, and $\ha$ (grey points).  A Spearman
      test on the individual measurements for the 93 galaxies
      indicates a correlation coefficient of $\rho = 0.54$, with a
      probability $p=2.3\times 10^{-8}$ that there is a null
      correlation between O32 and $\Sigma_{\rm SFR}$.  A linear fit to
      the data and the $1\sigma$ confidence interval are indicated by
      the blue line and shaded region, respectively: ${\rm O32} =
      (0.415 \pm 0.064)\log[\Sigma_{\rm SFR}/M_\odot\,{\rm
          yr}^{-1}\,{\rm kpc}^{-2}] + 0.307\pm 0.030$.  The average
      O32 values derived from composite spectra of galaxies in three
      bins of $\Sigma_{\rm SFR}$ are shown by the large green
      diamonds.}
    \label{fig:o32vssigma}
\end{figure}

Figure~\ref{fig:o32vssigma} shows the relationship between O32 and
$\Sigma_{\rm SFR}$ for 93 individual galaxies in the ``ionization
parameter'' sample (Section~\ref{sec:sample}) with significant (i.e.,
$S/N\ge 3$) detections of $\oii$, $\oiii$, $\hb$, and $\ha$.  Both
$\oii$ and $\oiii$ were corrected for dust attenuation by applying a
reddening correction derived from the Balmer decrement ($\ha/\hb$) and
assuming the \citet{cardelli89} extinction curve (e.g.,
\citealt{reddy20,rezaee21}).  The Balmer decrements for individual
galaxies vary from the theoretical value in a dust-free case, $\ha/\hb
\simeq 2.8$, up to $\ha/\hb \simeq 8$ for the dustiest objects in our
sample.  Also shown in this figure are the average O32 values derived
from composite spectra of galaxies in the ``ionization parameter''
sample in the same four $\Sigma_{\rm SFR}$ bins used to compute
$n_e$.\footnote{The average Balmer decrements measured in these four
  $\Sigma_{\rm SFR}$ bins, from lowest to highest $\Sigma_{\rm SFR}$,
  are $\langle\ha/\hb\rangle = 4.50\pm 0.40$, $4.32\pm 0.50$,
  $4.18\pm0.44$, and $3.70\pm 0.29$, respectively.  These composite
  Balmer decrement measurements are consistent with the mean Balmer
  decrement of individual galaxies (e.g., see \citealt{reddy15,
    shivaei15b}).}  These average O32 values imply that the
relationship between O32 and $\Sigma_{\rm SFR}$ for galaxies with
significant detections of the four aforementioned emission lines is
not substantially biased relative to that obtained for all 113
galaxies of the ``ionization parameter'' sample.

A Spearman test on the individual measurements indicates a significant
correlation between O32 and $\Sigma_{\rm SFR}$ for the subsample of 93
galaxies, with a probability $p=2.3\times 10^{-8}$ of a null
correlation between the two.  At face value, these results imply a
highly significant correlation between $U$ and $\Sigma_{\rm SFR}$.
However, it is important to assess the degree to which the translation
between O32 and $U$ may be affected by other parameters.  In
particular, a number of studies have investigated the effect of
ionizing spectral hardness, gas-phase abundance, and $n_e$ on the O32
ratio (e.g., \citealt{sanders16b, strom18}).  Here, we discuss some of
these dependencies in the context of updated stellar population
synthesis models that include the effects of stellar binarity (i.e.,
the binary BPASS models).  These models have generally been found to
simultaneously reproduce the rest-frame far-UV photospheric features
of high-redshift galaxies and their rest-frame optical nebular
emission line ratios (e.g., \citealt{steidel16, topping20a,
  topping20b, reddy22}).

\begin{figure}
  \epsscale{1.2}
  \plotone{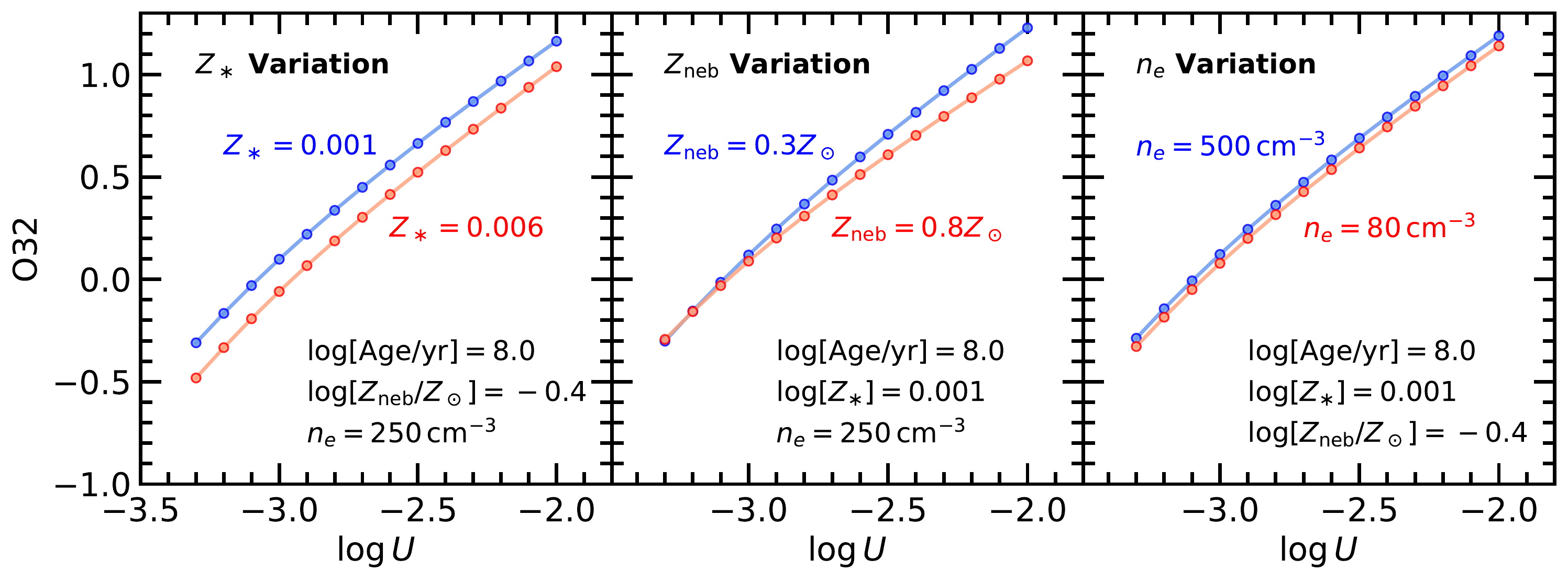}
    \caption{Predicted relationship between O32 and $\log U$ from
      photoionization modeling assuming a constant star formation
      BPASS stellar population model with the indicated stellar
      metallicity and age, and the indicated nebular oxygen abundance
      ($Z_{\rm neb}$) and $n_e$ (see text).}
    \label{fig:modelo32}
\end{figure}

Figure~\ref{fig:modelo32} shows the predicted O32 as a function of
$\log U$ for BPASS v2.2.1 constant star formation models with binary
stellar evolution and an upper-mass cutoff of the IMF of
$100$\,$M_\odot$ (i.e., the ``100bin'' models), with the input stellar
population and CLOUDY \citep{ferland17} parameters indicated in each
panel.\footnote{As shown elsewhere (e.g., \citealt{reddy12b,
    reddy22}), a constant star-formation history provides an adequate
  description of the average star-formation history for an ensemble of
  typical star-forming galaxies at $z\sim 2$, with
  stellar-population-derived ages consistent with what is expected
  given the dynamical timescale of these galaxies.}  For instance, the
left panel shows the relationship between O32 and $\log U$ for models
with a $\log[{\rm Age/yr}] = 8.0$ stellar population, an oxygen
abundance of $\log[Z_{\rm neb}/Z_\odot] = -0.4$\footnote{We assume
  throughout that $Z_\odot = 0.0142$ \citep{asplund09}.}, $n_e =
250$\,cm$^{-3}$---all values which are typical of MOSDEF galaxies
(e.g., \citealt{sanders16b, topping20a, reddy22})---and two stellar
metallicities ($Z_\ast = 0.001$ and $0.006$, expressed in terms of the
mass fraction of metals) that bracket the range inferred from modeling
the rest-frame FUV spectra of individual MOSDEF galaxies
\citep{topping20b}.  At a fixed O32, $\log U$ varies by $\simeq
0.15$\,dex for the aforementioned range of $Z_\ast$.  Similarly, the
middle panel shows the relationships between O32 and $\log U$ for two
values of $Z_{\rm neb}$ that bracket the range where the bulk of
MOSDEF galaxies lie (e.g., \citealt{topping20b}) and where the stellar
population age, $Z_\ast$, and $n_e$ are fixed to values typical of
MOSDEF galaxies.  Finally, the right panel indicates the relationships
between O32 and $\log U$ for two values of $n_e$ that bracket the
range inferred for the MOSDEF sample (\citealt{sanders16b}; see also
Section~\ref{sec:electrondensity}) and where all other parameters are
fixed to the typical values.  The model predictions summarized in
Figure~\ref{fig:modelo32} indicate that the translation between O32
and $\log U$ is relatively insensitive to $Z_\ast$, $Z_{\rm neb}$, and
$n_e$ over the ranges of these parameters that are represented in the
MOSDEF sample, changing at most by $\simeq 0.15$\,dex for a factor of
6 increase in $Z_\ast = 0.001$ to $0.006$.\footnote{Adopting an
  upper-mass cutoff of $300$\,$M_\odot$ for the IMF results in a
  relationship between O32 and $\log U$ that overlaps the one obtained
  for our fiducial model with an upper-mass cutoff of $100$\,$M_\odot$
  (e.g., \citealt{steidel16}).}  Thus, the $\simeq 0.8$\,dex increase
in O32 over the range of $\Sigma_{\rm SFR}$ shown in
Figure~\ref{fig:o32vssigma} likely reflects an increase in $U$ with
$\Sigma_{\rm SFR}$.

\begin{figure}
  \epsscale{1.2}
  \plotone{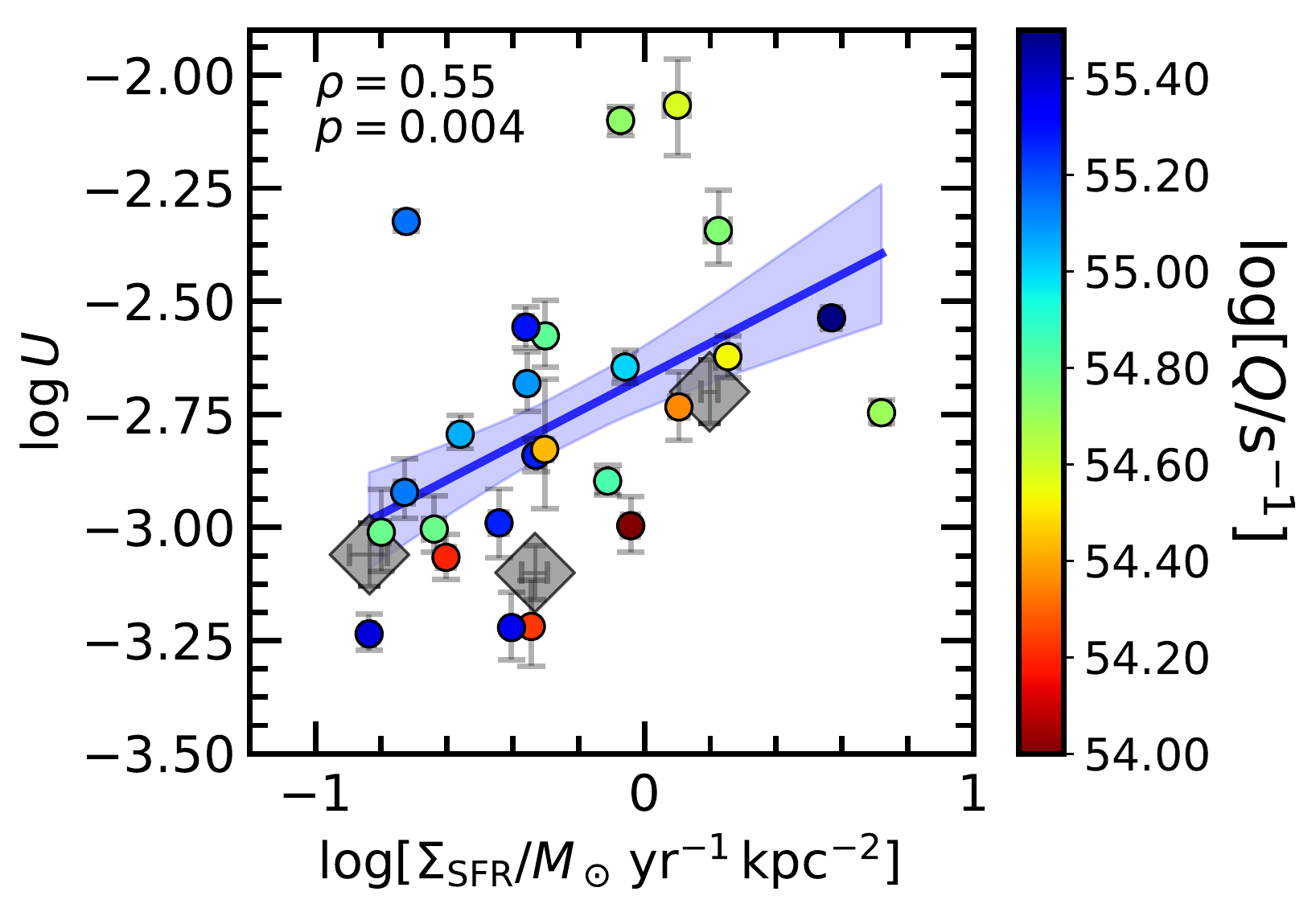}
    \caption{Relationship between $\log U$ and $\log[\Sigma_{\rm
          SFR}/M_\odot\,{\rm yr}^{-1}\,{\rm kpc}^{-2}]$ for 25
      individual galaxies in the MOSDEF-LRIS sample \citep{topping20b}
      with robust size, and hence $\Sigma_{\rm SFR}$, measurements
      (circles), and detections of $\ha$, color coded by $\log[Q/{\rm
          s}^{-1}]$ (see Section~\ref{sec:q}) .  The Spearman
      correlation coefficient and $p$-value are indicated.  The solid
      blue line and shaded region indicate a linear fit to the data
      and the $1\sigma$ confidence interval, respectively, with the
      following relation: $\log U = (0.379\pm
      0.030)\log\left[\frac{\Sigma_{\rm SFR}}{M_\odot\,{\rm
            yr}^{-1}\,{\rm kpc}^{-2}}\right] - 2.668\pm0.068$.  .
      Average values obtained from fitting composite spectra in three
      equal-number bins of $\Sigma_{\rm SFR}$ (from \citealt{reddy22})
      are shown by the large grey diamonds.}
    \label{fig:loguvssigma}
\end{figure}

This conclusion is further corroborated by direct modeling of a subset
of MOSDEF galaxies with deep rest-frame FUV spectra (i.e., the
MOSDEF-LRIS sample; \citealt{topping20a, reddy22}).  \citet{reddy22}
self-consistently modeled the rest-frame FUV spectra and rest-frame
optical emission line ratios of galaxies in the MOSDEF-LRIS sample for
galaxies in each of three equal-number bins of $\Sigma_{\rm SFR}$.
Galaxies in the lower, middle, and upper third of the $\Sigma_{\rm
  SFR}$ distribution have $\langle {\rm O32}\rangle = -0.03\pm 0.14$,
$0.03\pm 0.09$, and $0.33\pm 0.07$, respectively.  The modeling of the
composite rest-frame FUV spectra of galaxies in these three bins of
$\Sigma_{\rm SFR}$ indicates $\langle Z_\ast\rangle = 0.0013 \pm
0.0005$, $0.0021\pm 0.0008$, and $0.0017\pm 0.0004$, respectively (see
Table~3 of \citealt{reddy22}).  Based on the left panel of
Figure~\ref{fig:modelo32}, this variation in $\langle Z_\ast\rangle$
results in a negligible shift in the relationship between O32 and
$\log U$.  Similarly, $\langle Z_{\rm neb}/Z_\odot \rangle = 0.28\pm
0.13$, $0.40\pm 0.03$, and $0.43\pm 0.08$, respectively, for the three
aforementioned bins of $\Sigma_{\rm SFR}$ (Table~3 of
\citealt{reddy22}).  Based on the middle panel of
Figure~\ref{fig:modelo32}, this variation in $\langle Z_{\rm
  neb}\rangle$ also implies a negligible shift in the relationship
between O32 and $\log U$.  Finally, as per the discussion in
Section~\ref{sec:electrondensity}, though there is a significant
correlation between $n_{e}$ and $\Sigma_{\rm SFR}$, the range of
inferred $n_e$ implies a negligible shift in the translation between
O32 and $\log U$.  In summary, the variations in $\langle
Z_\ast\rangle$, $\langle Z_{\rm neb}\rangle$, and $\langle n_e\rangle$
for composites constructed in bins of $\Sigma_{\rm SFR}$ are unlikely
to be solely responsible for driving the observed increase in O32 with
$\Sigma_{\rm SFR}$ (Figure~\ref{fig:o32vssigma}).  Rather, this
relationship is likely driven by changes in $U$.  To
that point, \citet{reddy22} calculated $\langle \log U\rangle =
-3.06\pm 0.07$, $-3.10\pm 0.06$, and $-2.70\pm 0.07$, respectively,
for the three bins of $\Sigma_{\rm SFR}$
(Figure~\ref{fig:loguvssigma}), implying a significantly higher
$\langle \log U\rangle $ for galaxies in the upper third of the
$\Sigma_{\rm SFR}$ distribution.

Finally, the relationship between $\log U$ and $\Sigma_{\rm SFR}$ for
a subset of 25 individual MOSDEF-LRIS galaxies from \citet{topping20b}
with robust size measurements and $\lha/\sigma_{\lha}\ge 3$, where
$\lha$ is the dust-corrected $\ha$ luminosity, is shown in
Figure~\ref{fig:loguvssigma}.\footnote{We imposed a requirement on the
  significance of the $\ha$ luminosity in order to compute the
  ionizing photon rate ($Q$) for these galaxies, as discussed in
  Section~\ref{sec:q}.}  A Spearman correlation test indicates a
probability of $p=0.004$ that the two variables are uncorrelated,
suggesting that the correlation is significant at the $\approx
3\sigma$ level.  The figure also shows a linear fit to the data for
the individual galaxies.

In summary, the significant correlation between O32 and $\Sigma_{\rm
  SFR}$ (Figure~\ref{fig:o32vssigma}); the insensitivity of the
translation between O32 and $U$ to $Z_\ast$, $Z_{\rm neb}$, and $n_e$
over the range of these parameters represented in the MOSDEF sample
(Figure~\ref{fig:modelo32}); and the differences in average and
individually-inferred $U$ for galaxies with low and high $\Sigma_{\rm
  SFR}$ (Figure~\ref{fig:loguvssigma}) altogether suggest a genuine
correlation between $U$ and $\Sigma_{\rm SFR}$ (see also
\citealt{bian16, runco21, reddy22}).

\section{\bf DISCUSSION}
\label{sec:discussion}

In this section, we examine the results on $n_e$ and $U$ in the
context of the factors that $U$ depends on, including the ionizing
photon rate (Section~\ref{sec:q}), $n_e$ (Section~\ref{sec:ne}), the
volume filling factor of dense clumps in $\hii$ regions
(Section~\ref{sec:epsilon}), and the escape fraction of ionizing
photons (Section~\ref{sec:fesc}).  The implications of our results for
the redshift evolution of $U$ are discussed in
Section~\ref{sec:evolution}.

\subsection{Key Factors that Modulate the Ionization Parameter of High-Redshift Galaxies}
\label{sec:keyfactors}

Several previous efforts have focused on understanding the factors
responsible for the elevated ionization parameters inferred for
high-redshift galaxies (e.g., \citealt{brinchmann08, bian16}).  Here,
we extend upon these previous works by concentrating on the factors
responsible for the correlation between $U$ and $\Sigma_{\rm SFR}$
among high-redshift galaxies, taking advantage of the most up-to-date
inferences of $n_e$, ionizing photon rates ($Q$), volume filling
factors ($\epsilon$), and the escape fraction of ionizing photons
($f_{\rm esc}$) for the same galaxies.  Equation~\ref{eq:U} summarizes
the dependencies between $U$ and $Q$, $n_e$, and $\epsilon$.  Below,
we discuss each of these factors, along with $f_{\rm esc}$, in turn.

\begin{figure}
  \epsscale{1.2}
  \plotone{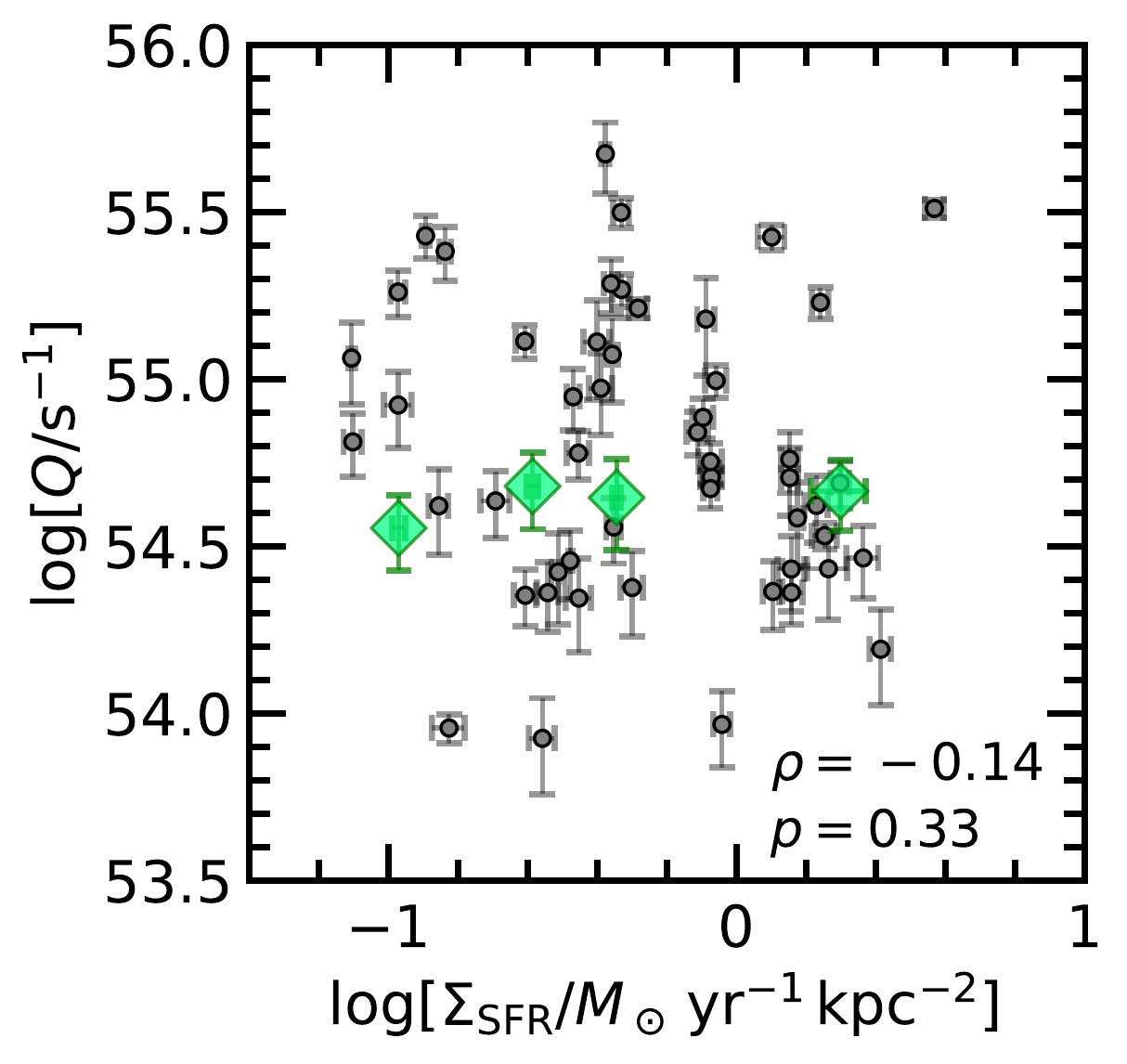}
    \caption{Relationship between $\log [Q/{\rm s}^{-1}]$ and
      $\log[\Sigma_{\rm SFR}/M_\odot\,{\rm yr}^{-1}\,{\rm kpc}^{-2}]$
      for the 51 galaxies in the ``ionization parameter'' sample with
      $Q/\sigma_Q\ge 3$ (grey points).  The Spearman correlation
      coefficient and $p$-value are indicated.  Average values
      obtained from fitting composite spectra of galaxies in the
      ``ionization parameter'' sample in the same four bins of
      $\Sigma_{\rm SFR}$ used to compute $n_e$ are shown by the large
      green diamonds.}
    \label{fig:logqvslogu}
\end{figure}

\subsubsection{Ionizing Photon Rates ($Q$)}
\label{sec:q}

Figure~\ref{fig:logqvslogu} shows the variation in $Q$ with
$\Sigma_{\rm SFR}$ for the 51 galaxies in the ``ionization parameter''
sample with $Q/\sigma_Q \ge 3$, where $\sigma_Q$ is the uncertainty in
$Q$.\footnote{The uncertainty in $Q$, $\sigma_Q$, includes the
  uncertainty in the Balmer-decrement-inferred dust correction to the
  $\ha$ luminosity.}  The \citet{leitherer95} relation was used to
convert dust-corrected $\lha$ to $Q$.\footnote{We did not apply any
  upward correction to $Q$ to account for the escape fraction of
  ionizing photons, $f_{\rm esc}$.  Doing so systematically shifts
  $\log[Q/{\rm s}^{-1}]$ higher by $\la 0.04$\,dex and does not affect
  any of our conclusions.}  Also shown are the average values obtained
from composite spectra of the 113 galaxies in the ``ionization
parameter'' sample in the same four bins of $\Sigma_{\rm SFR}$ used to
compute $n_e$.  A Spearman correlation test on the individual
measurements of $Q$ and $\Sigma_{\rm SFR}$ imply that the two are not
significantly correlated, a result confirmed by the invariance of
$\langle Q\rangle$ for the four $\Sigma_{\rm SFR}$ bins.

The ionizing photon rate depends on the SFR and the ionizing photon
production efficiency, $\xi_{\rm ion}$ \citep{robertson13, bouwens16a,
  shivaei18, theios19, reddy22}: $Q\propto \xi_{\rm ion}\times {\rm
  SFR}$.  The ionizing photon production efficiency depends on the
specific details of the massive stars, including their stellar
metallicities, ages, whether they evolve as single stars or in
binaries, and the IMF.  \citet{reddy22} showed that $\xi_{\rm ion}$
does not vary significantly with $\Sigma_{\rm SFR}$ for galaxies in
the MOSDEF-LRIS sample.  This result, combined with the lack of a
strong correlation between $\lha$ and $\Sigma_{\rm SFR}$ (note that
the latter is based on SED-inferred SFRs; Section~\ref{sec:sample}),
results in an average $Q$ that is invariant over the dynamic range of
$\Sigma_{\rm SFR}$ probed by our sample.  Thus, the increase in $U$
with $\Sigma_{\rm SFR}$ cannot be explained by changes in $Q$ alone.

\subsubsection{Electron Densities ($n_e$)}
\label{sec:ne}

As noted in Section~\ref{sec:electrondensity} and shown in
Figure~\ref{fig:density2}, we find a significant correlation between
$\langle n_e\rangle$ and $\langle \Sigma_{\rm SFR}\rangle$: galaxies
with $\langle\Sigma_{\rm SFR}\rangle \simeq
2$\,$M_\odot$\,yr$^{-1}$\,kpc$^{-2}$ have $\langle n_e\rangle \simeq
400$\,cm$^{-3}$, a factor of $\approx 5$ larger than that of galaxies
with $\langle\Sigma_{\rm SFR}\rangle \simeq
0.1$\,$M_\odot$\,yr$^{-1}$\,kpc$^{-2}$.  For fixed $Q$ and $\epsilon$,
this increase in $n_e$ corresponds to an $\approx 0.23$\,dex increase
in $\log U$ based on Equation~\ref{eq:U}, and accounts for roughly
half of the observed $0.4-0.5$\,dex increase in $\log U$ over the
aforementioned range of $\Sigma_{\rm SFR}$
(Figure~\ref{fig:loguvssigma}).\footnote{Note that $U$ is primarily
  constrained by the O32 index, while $Q$ and $\epsilon$ depend on the
  dust-corrected $\ha$ luminosity and $n_e$ is constrained by the
  ratio of $\oii$\,$\lambda 3730$ to $\oii$\,$\lambda 3727$.  As such,
  aside from some common dependence on the Balmer-decrement-determined
  dust correction used to compute $Q$ and $\epsilon$, $U$ is sensitive
  to a combination of emission lines that is relatively independent of
  those used to constrain $Q$, $n_e$, and $\epsilon$.}  Thus, the
dependence of $n_e$ on $\Sigma_{\rm SFR}$ is an important contributing
factor to the dependence of $\log U$ on $\Sigma_{\rm SFR}$.

Note that $n_e$ and $\Sigma_{\rm SFR}$ are sensitive to gas on
different physical scales.  Specifically, $\Sigma_{\rm SFR}$ in the
present analysis probes star formation on galactic-wide (kpc) scales,
while $n_e$ is sensitive to dense structures within pc-scale $\hii$
regions.  A simple explanation for why the two may correlate is that
the Kennicutt-Schmidt relation \citep{kennicutt98} connects
$\Sigma_{\rm SFR}$ to the molecular gas density, and the latter, along
with the external ambient density and/or interstellar pressure,
determines $n_e$ of $\hii$ regions (\citealt{shirazi14, shimakawa15,
  kashino19, jiang19}; see further discussion in \citealt{davies21}).
Higher spatial resolution measurements of molecular gas and electron
densities afforded by nearby galaxies and AO-assisted observations of
unlensed and/or lensed galaxies at high redshift should further
elucidate the connection between $\Sigma_{\rm SFR}$ and $n_e$.

\subsubsection{Volume Filling Factors ($\epsilon$)}
\label{sec:epsilon}

The variable $\epsilon$ represents the volume filling factor of line-emitting
structures within the otherwise diffuse ionized gas in $\hii$ regions, and
can be approximated by 
\begin{equation}
\epsilon \approx [n_{e,{\rm rms}} / n_e]^2,
\label{eq:epsilon}
\end{equation}
where $n_e$ is the average density of gas giving rise to the $\oii$
emission and $n_{e,{\rm rms}}$ is the rms electron density in $\hii$
regions \citep{osterbrock59, kennicutt84}.  The latter depends on the
volume of $\hii$ regions which cannot be directly constrained as
individual $\hii$ regions are unresolved by our observations.  If we
assume that $\hii$ regions fill the entire star-forming volume, then
a lower limit on $n_{e,{\rm rms}}$ can be calculated from the dust-corrected
$\ha$ luminosity and the (e.g., spherical) volume of
ionized gas within a half-light radius, $V_{\hii}(<R_{\rm eff})$:
\begin{equation}
n_{e,{\rm rms}} = \left[\frac{\lha}{2\gamma_{\ha}V_{\hii}(<R_{\rm eff})}\right]^{1/2},
\label{eq:nerms}
\end{equation}
where $\gamma_{\ha}$ is the volume emissivity of $\ha$ (e.g.,
\citealt{rozas96, davies21}).  For Case B recombination and $T_e =
10^4$\,K, $\gamma_{\ha} = 3.56\times
10^{-25}$\,erg\,cm$^{3}$\,s$^{-1}$.  The factor of $2$ in the
denominator of Equation~\ref{eq:nerms} accounts for the fact that the
volume is calculated based on the half-light radius.

Equations~\ref{eq:epsilon} and \ref{eq:nerms} were used to compute
lower limits on $\langle n_{e,{\rm rms}}\rangle$ and
$\langle\epsilon\rangle$ from the average $\ha$ luminosities and sizes
of galaxies in the ``ionization parameter'' sample in the same four
$\Sigma_{\rm SFR}$ bins used to compute $n_e$
(Section~\ref{sec:electrondensity}).  These lower limits span the
range $\langle n_{e,{\rm rms}}\rangle \simeq 1$ to $6$\,cm$^{-3}$ and
$\langle \epsilon\rangle\simeq 10^{-4} - 10^{-3}$.  These values are
broadly consistent with the average $n_{e,{\rm rms}}$ and $\epsilon$
found for star-forming galaxies at similar redshifts from the
KMOS$^{\rm 3D}$ survey \citep{davies21}.  Evidently the gas that
dominates the line emission likely constitutes a small fraction of the
total ionized volume in high-redshift galaxies (see also
\citealt{kennicutt84, rozas96, elmegreen00, copetti00, hunt09,
  davies21}).  At any rate, given that the $n_{e,{\rm rms}}$ and
$\epsilon$ calculated above represent lower limits, we cannot rule out
the possibility that there may be a correlation between $\epsilon$ and
$\Sigma_{\rm SFR}$ (see Section~\ref{sec:summary} for further
discussion).

\subsubsection{Ionizing Escape Fractions ($f_{\rm esc}$)}
\label{sec:fesc}

Line diagnostics that are typically used to infer $U$, such as O32,
will overestimate $U$ if there is a non-zero escape fraction of
ionizing photons, $f_{\rm esc}$ \citep{giammanco05, brinchmann08,
  nakajima13}.  A simple example is the case of a density-bounded
nebula where the region of low-ionization emission (e.g., $\oii$) is
truncated, leading to higher O32 and higher apparent $U$.  We can
evaluate the magnitude of this effect using recent determinations of
$f_{\rm esc}$ at high redshift.  

In particular, \citet{reddy22} inferred typical average escape
fractions of $\langle f_{\rm esc}\rangle \la 10\%$ based on the
average depths of Lyman series absorption lines in the composite
spectra of galaxies in the MOSDEF-LRIS sample (see also
\citealt{reddy16b}), with no significant difference in the inferred
$\langle f_{\rm esc}\rangle$ for galaxies in the lower- and
upper-third of the distribution of $\Sigma_{\rm SFR}$ (e.g., Figure~18
in \citealt{reddy22}).  The typical $\langle f_{\rm esc}\rangle$ found
in that study is similar to the sample-averaged values of $\langle
f_{\rm esc}\rangle \simeq 6-9\%$ derived by \citet{steidel18} and
\citet{pahl21} for a sample of typical star-forming galaxies at $z\sim
3$.  Based on the modeling of \citet{giammanco05}, this low value of
$f_{\rm esc}$ results in an apparent $\log U$ that is $\la 0.2$\,dex
higher than the value when $f_{\rm esc}=0$.\footnote{Virtually all of
  the galaxies in our sample have O32$\la 0.7$, significantly lower
  than the values (O32$\ga 1$) typically associated with
  density-bounded $\hii$ regions with purportedly high ionizing escape
  fractions (e.g., \citealt{jaskot13, nakajima13}).}  This small
change in inferred $\log U$, combined with the lack of a significant
correlation between $\langle f_{\rm esc}\rangle$ and $\langle
\Sigma_{\rm SFR}\rangle$ for galaxies in the MOSDEF-LRIS sample,
suggests that the trend between $\log U$ and $\Sigma_{\rm SFR}$ is
unlikely to be related to variations in $f_{\rm esc}$.

\subsubsection{Summary}
\label{sec:summary}

The results of the previous sections can be summarized as follows.  We
find significant correlations between $n_e$ and $\Sigma_{\rm SFR}$
(Section~\ref{sec:electrondensity}) and between $U$ and $\Sigma_{\rm
  SFR}$ (Section~\ref{sec:ionizationparameter}) for typical
star-forming galaxies at $z\sim 2$.  Thus, $n_e$ appears to be an
important factor in explaining the correlation between $U$ and
$\Sigma_{\rm SFR}$ (Section~\ref{sec:ne}).  Of the factors that $U$
depends on, $\epsilon$ is perhaps the most uncertain since it requires
knowledge of the volumes of spatially-unresolved $\hii$ regions, and
we cannot rule out the possibility that $\epsilon$ may also play a
role in shaping the correlation between $U$ and $\Sigma_{\rm SFR}$
(Section~\ref{sec:epsilon}).  Indeed, it is not unreasonable to expect
that the volume filling factor of $\hii$ regions increases with
$\Sigma_{\rm SFR}$, in which case $\epsilon$ may play a significant
role in driving the relationship between $U$ and $\Sigma_{\rm SFR}$.
On the other hand, there is no significant correlation between $Q$ and
$\Sigma_{\rm SFR}$ for galaxies in our sample (Section~\ref{sec:q}),
suggesting that changes in the ionizing photon rate are unlikely to
contribute to the observed relationship between $U$ and $\Sigma_{\rm
  SFR}$.  Similarly, $f_{\rm esc}$ inferred for galaxies in the sample
are too low to account for the significant correlation between $U$ and
$\Sigma_{\rm SFR}$ (Section~\ref{sec:fesc}).

While $n_e$ appears to be an important factor in driving the
relationship between $U$ and $\Sigma_{\rm SFR}$ at $z\sim 2$, it is
clear that parameters other than $n_e$ may be important for explaining
the scatter in this relationship.  For instance,
Figure~\ref{fig:loguvssigma} shows that the five galaxies with the
lowest measured $Q$ in the \citet{topping20b} sample generally lie
below the mean relation between $U$ and $\Sigma_{\rm SFR}$.  Thus,
variations in $Q$ may be partly responsible for the scatter in the
relationship between $U$ and $\Sigma_{\rm SFR}$.
Figure~\ref{fig:loguvssigma} also shows that there are some galaxies
with similar $\Sigma_{\rm SFR}$ and $Q$, but which have significantly
different $U$.  This result suggests that variations in $\epsilon$ may
also be important for explaining some of the scatter in the
relationship between $U$ and $\Sigma_{\rm SFR}$.

\subsection{Redshift Evolution of $U$}
\label{sec:evolution}

So far we have focused on the factors that drive the relationship
between $U$ and $\Sigma_{\rm SFR}$ for $z\sim 2$ galaxies.  We can
further explore the extent to which the relationship between $U$ and
$\Sigma_{\rm SFR}$ (or $n_e$) contributes to the redshift evolution of
$U$, as noted by a number of recent studies (Section~\ref{sec:intro}).
For example, \citet{sanders16b} noted that typical star-forming
galaxies at $z\sim 2$ from the MOSDEF survey have O32 that are on
average $\simeq 0.6$\,dex higher than local galaxies at a fixed
stellar mass.  This offset in O32 suggests that $z\sim 2$ galaxies
have on average a higher ionization parameter relative to local
galaxies at a fixed stellar mass.

In particular, the average O32 of $\sim 10^{10}$\,$M_\odot$ galaxies
at $z\sim 2$ is $\langle {\rm O32}\rangle \simeq 0.1$
\citep{sanders16b}.  Based on the model predictions shown in
Figure~\ref{fig:modelo32}, this average O32 corresponds to $\log U
\simeq -3.0$ (see also \citealt{topping20a, runco21, reddy22}).  Local
star-forming galaxies of the same stellar mass from the SDSS sample
have $\langle {\rm O32}\rangle \simeq -0.6$, corresponding to $\log U
\simeq -3.6$ \citep{shirazi14, nakajima14}.  Thus, at fixed stellar
mass of $\sim 10^{10}$\,$M_\odot$, galaxies at $z\sim 2$ have $\langle
\log U\rangle$ that is $\approx 0.6$\,dex higher than that of local
galaxies.  Below, we discuss some of the factors that may be
responsible for the redshift evolution of $U$.

\subsubsection{The Role of Electron Density and SFR}
\label{sec:neandsfr}

\citet{sanders16b} found an order of magnitude increase in $\langle
n_e\rangle$ for galaxies with stellar masses of $\sim
10^{10}$\,$M_\odot$ from $z\sim 0$ to $z\sim 2$.  In the simple
spherical geometry of an ionization-bounded nebula where $U\propto
n_e^{1/3}$ (Equation~\ref{eq:U}), such an increase in $n_e$ translates
to an $\approx 0.3$\,dex increase in $\log U$ for a fixed $Q$ and
$\epsilon$.  Thus, at face value, the redshift evolution of $n_e$
could account for a significant fraction of the $\simeq 0.6$\,dex
increase in $\log U$ for $z\sim 2$ galaxies relative to local galaxies
at fixed stellar mass \citep{brinchmann08, shirazi14, bian16}.

In addition, there is an $\approx 1$\,dex increase in SFR (and hence
$Q$, assuming a constant $\xi_{\rm ion}$) between $z\sim 0$ and $z\sim
2$ at a fixed stellar mass of $10^{10}$\,$M_\odot$ (e.g.,
\citealt{speagle14}).  This redshift evolution of SFR (and $Q$)
implies an $\approx 0.3$\,dex increase in $\log U$ assuming the
scaling relation specified in Equation~\ref{eq:U} (see also
\citealt{nakajima14, kaasinen18}).  Finally, \citet{davies21} present
evidence that the volume filling factor, $\epsilon$, does not evolve
significantly over the redshift range $0\la z\la 2.6$.  Hence, the
$0.3$\,dex increase in $\log U$ due to $n_e$ evolution, and the
$0.3$\,dex increase in $\log U$ due to SFR evolution, together could
account for much of the $0.6$\,dex increase in $\log U$ from $z\sim 0$
to $z\sim 2$.  In the next section, we evaluate this conclusion in the
context of previous studies that have underscored the role of
metallicity in the redshift evolution of $U$.

\subsubsection{The Role of Stellar Metallicity}
\label{sec:metallicity}

Several studies have attributed the redshift evolution of $U$ at a
fixed stellar mass to changes in metallicity.  Specifically, there is
a well-established anti-correlation between $U$ and $Z_{\rm neb}$ for
local star-forming galaxies (e.g., \citealt{dopita86, dopita06,
  perez14}) which is usually explained in terms of lower-metallicity
massive stars having harder ionizing spectra and more intense
radiation fields \citep{dopita06, leitherer14}.  In this case,
$\xi_{\rm ion}$, and hence $Q$ at a fixed SFR, will be larger for
lower-metallicity stellar populations.  The redshift-invariance of the
relationship between $U$ and $Z_{\rm neb}$ \citep{topping20b,
  sanders20} then implies that the decrease in $Z_{\rm neb}$ with
redshift at a fixed stellar mass (i.e., the redshift evolution of the
mass-metallicity relation) is accompanied by an increase in $U$ with
redshift at a fixed stellar mass (e.g., \citealt{sanders16b}).  In the
following discussion, we examine the extent to which metallicity
affects the redshift evolution of $U$.

\begin{figure}
  \epsscale{1.2}
  \plotone{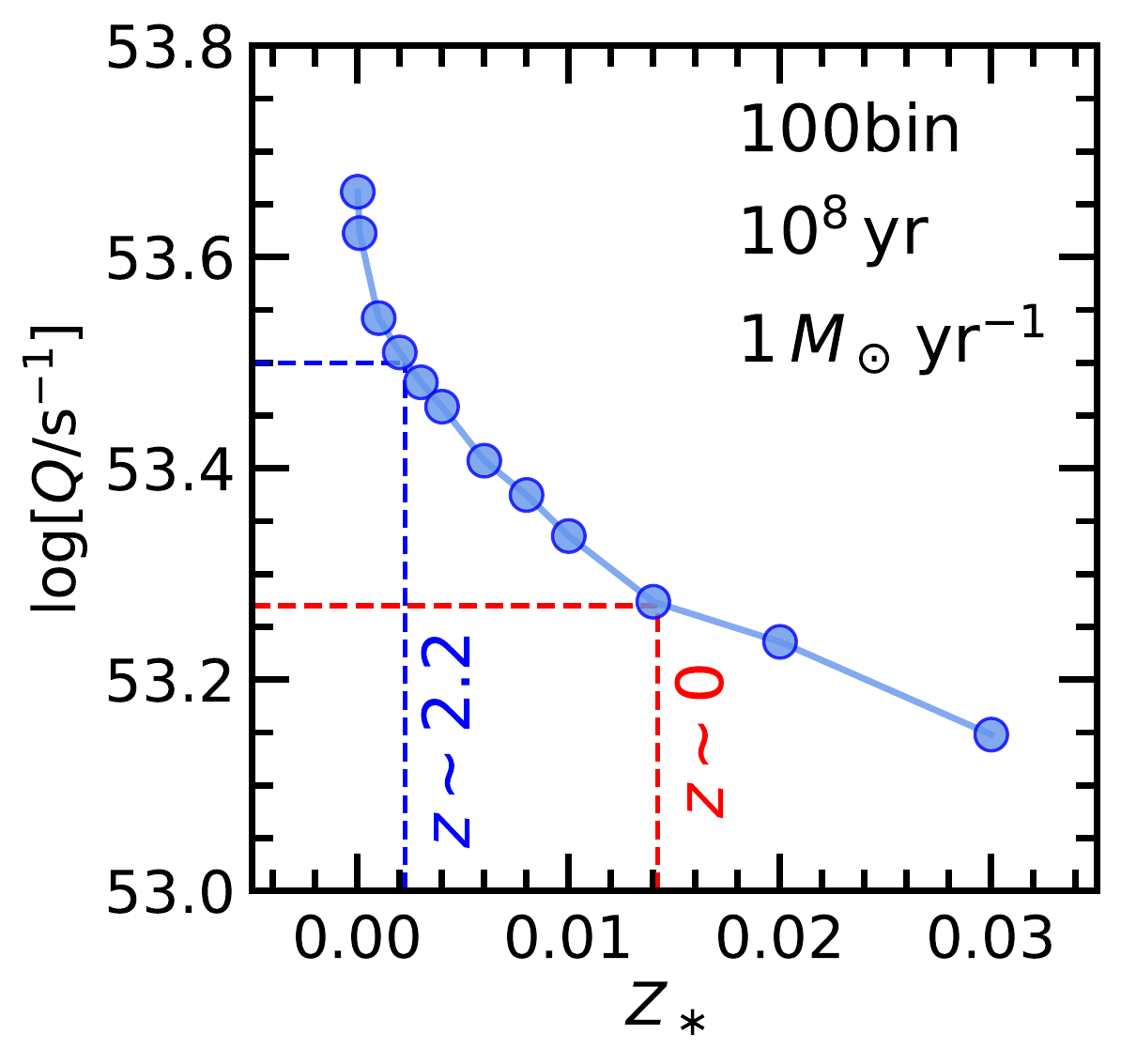}
    \caption{$Q$ as function of $Z_\ast$ for the BPASS 100bin models
      with an age of $10^8$\,yr and a constant star-formation rate of
      $1$\,$M_\odot$\,yr$^{-1}$.  Also indicated are the $\langle
      Z_\ast\rangle$ inferred for $z\sim 0$ and $z\sim 2.2$ galaxies
      at a fixed stellar mass of $\sim 10^{10}$\,$M_\odot$ from
      \citet{kashino22}: $\langle Z_\ast\rangle \simeq 0.0142$
      (roughly solar) and $\langle Z_\ast\rangle \simeq 0.002$,
      respectively.  The model predictions indicate $\langle
      \log[Q/{\rm s}^{-1}]\rangle \simeq 53.50$ and 53.27,
      respectively, for the these two values of $\langle
      Z_\ast\rangle$.}
    \label{fig:qvszstar}
\end{figure}

The ionizing spectra of massive stars are more directly connected to
stellar metallicity (or Fe abundance) than O abundance ($Z_{\rm
  neb}$), since the former dominates the opacity of stellar
atmospheres and regulates the launching of stellar winds and the
absorption of ionizing photons by those winds (e.g.,
\citealt{dopita06}).  As such, we frame our discussion in terms of
$Z_\ast$ rather than $Z_{\rm neb}$\footnote{In general, $Z_\ast$ has
  been found to lag $Z_{\rm neb}$ for $z\ga 2$ galaxies, an effect
  that has been attributed to $\alpha$-enhanced stellar populations at
  these redshifts \citep{steidel16, cullen19, topping20a, cullen21,
    reddy22}.\label{foot:alpha}}.  Figure~\ref{fig:qvszstar} shows how
$Q$ is predicted to vary with $Z_\ast$ for the BPASS 100bin stellar
population synthesis models with an age of $10^8$\,yr and a constant
star-formation rate of $1$\,$M_\odot$\,yr$^{-1}$.  Also indicated are
$\langle Z_\ast\rangle$ inferred for $z\sim 0$ and $z\sim 2.2$
galaxies with $M_\ast \sim 10^{10}$\,$M_\odot$ based on the
relationships between stellar metallicity and stellar mass at those
redshifts \citep{kashino22}.  Based on the model predictions, the
difference in $\langle Z_\ast\rangle$ inferred at these two redshifts
results in $\Delta\log[Q/{\rm s}^{-1}] \simeq 0.23$\,dex.  If all
other parameters affecting $Q$ are held fixed (i.e., the details of
the stellar population model including the star-formation history,
age, inclusion of binaries, IMF, and SFR), and if $n_e$ and $\epsilon$
are held fixed, then the difference in $Z_\ast$ between $z\sim 0$ and
$z\sim 2.2$ galaxies implies $\Delta\log U \simeq 0.08$\,dex assuming
$U\propto Q^{1/3}$ (Equation~\ref{eq:U}).

Note that the stellar population characteristics (star-formation,
history, age, etc.) of $10^{10}$\,$M_\odot$ galaxies at $z\sim 0$ may
be different that those of similar-mass galaxies at $z\sim 2$.
Varying these other properties can also influence $Q$.  However, our
goal here is to determine the effects of metallicity {\em alone} on
$Q$, keeping all other parameters fixed.  In that case, the
$0.08$\,dex change in $\log U$ due to metallicity effects alone is
smaller than the $0.3$\,dex change in $\log U$ that can be attributed
to either the evolution of $n_e$ or SFR with redshift at a fixed
stellar mass (Section~\ref{sec:neandsfr}).

Note also that there are non-negligible systematic uncertainties in
$Z_\ast$ due to SPS-model variations in the predicted strengths of
stellar photospheric lines at a fixed $Z_\ast$, and the specific
wavelength ranges used to fit these models to observed spectra at
different redshifts (e.g., \citealt{cullen19, kashino22}).  However,
even in the extreme comparison of stellar populations with primordial
and super-solar abundances, the models predict $\Delta\log[Q/{\rm
    s}^{-1}] \simeq 0.51$\,dex, corresponding to $\Delta\log U\simeq
0.17$\,dex, which is still smaller than the changes in $U$ induced by
$n_e$ or SFR evolution.  Thus, if we assume the scaling relations
specified by Equation~\ref{eq:U} hold for both local and high-redshift
$\hii$ regions, then our results suggest that the redshift evolution
of $U$ at a fixed stellar mass is primarily due to variations in $n_e$
and SFR, with metallicity being a subdominant factor.

On the other hand, \citet{sanders16b} suggest that metallicity is the
primary factor driving the evolution in $U$ at a fixed stellar mass.
They point out that an increase in $n_e$ could be compensated by a
decrease in $\epsilon$, resulting in $U$ that is dominated by
variations in $Q$.  For this to occur, $\epsilon$ would have to
decrease with increasing redshift, which is at odds with the apparent
lack of redshift evolution of $\langle\epsilon\rangle$
\citep{davies21}.  Aside from the joint evolution of $n_e$ and
$\epsilon$, one would also have to account for the impact of the
evolving SFR (at a fixed stellar mass) on $Q$.

There are two additional points to consider.  First, our conclusions
regarding the importance of $n_e$ and SFR on the redshift evolution of
$U$ rely on the scaling relations of Equation~\ref{eq:U}.  These
simple relations may not apply to real $\hii$ regions with geometries
that depart from that of a simple ionization-bounded Str\"{o}mgren
sphere (see discussion in \citealt{sanders16b}).  Recall that $Q
\propto \xi_{\rm ion}\times {\rm SFR}$ (Section~\ref{sec:q}).  For
metallicity to be a dominant factor in the redshift evolution of $U$,
one would have to conceive of a scenario where $U$ scales more
strongly with $\xi_{\rm ion}$ than with SFR, which seems unlikely
given that $U$ depends on the {\em total} ionizing photon rate, $Q$;
or that $U$ scales strongly with $\xi_{\rm ion}$ and scales weakly
with some combination of SFR, $n_e$, and $\epsilon$.  It is unclear
what $\hii$ region geometries or ISM states could satisfy these
conditions.

Second, $U$ may depend more strongly on stellar metallicity if the
shape of the ionizing spectrum at a fixed stellar metallicity becomes
harder with increasing redshift (i.e., if the relationship between
$Q/{\rm SFR}$, or $\xi_{\rm ion}$, and $Z_\ast$ shown in
Figure~\ref{fig:qvszstar} evolves with redshift).  However, this
scenario would pose a problem for self-consistently explaining both
the stellar metallicities and the ionizing photons rates inferred for
$z\sim 2$ galaxies.  Instead, the BPASS models that best fit the
rest-frame far-UV stellar photospheric features (which determines
$Z_\ast$) also predict SFRs based on the total ionizing photon rate
(e.g., $\ha$-based SFRs) that are consistent with those derived from
the non-ionizing UV continuum \citep{reddy22}.

\subsubsection{Concluding Remarks}

Given the above discussion, we favor the simplest explanation for the
redshift evolution of $U$ at a fixed stellar mass; i.e., one in which
this evolution is driven by the order of magnitude increases in $n_e$
and SFR from $z\sim 0$ to $z\sim 2$ at a fixed stellar mass.  This
result does not necessarily conflict with the finding that galaxies at
a fixed nebular abundance (O/H) have similar $U$ irrespective of
redshift (e.g., \citealt{topping20b, sanders20}), or that $U$ strongly
anti-correlates with $Z_{\rm neb}$ at $z\sim 0$ for the reasons given
in \citet{strom18}.  The anti-correlation between $U$ and $Z_{\rm
  neb}$ does not necessarily imply that $Z_{\rm neb}$ is the {\em
  causative} factor in explaining the redshift evolution in $U$ at a
fixed stellar mass.  Rather, we suggest that there are other factors
that anti-correlate with $Z_{\rm neb}$ (i.e., gas density) that are
responsible for much of the redshift evolution in $U$ at a fixed
stellar mass.

Here, we simply point out that local galaxies with the same $Z_\ast$
as $10^{10}$\,$M_\odot$ galaxies at $z\sim 2$ are inferred to have
stellar masses that are at least two orders of magnitude lower (i.e.,
$M_\ast\la 10^{8}$\,$M_\odot$) based on the local and $z\sim 2.2$
relations between stellar metallicity and stellar mass.  These
low-mass local galaxies generally exhibit higher specific SFRs (sSFRS)
than more massive local star-forming galaxies (e.g.,
\citealt{laralopez10,cook14}), with the former being similar to the
sSFRs of typical star-forming galaxies at $z\sim 2$.  Furthermore,
there is tentative evidence for a significant ($4\sigma$) correlation
between sSFR and $n_e$ at $z\sim 2$ \citep{shimakawa15}.\footnote{We
  cannot independently confirm the presence of such a correlation at
  $z\sim 2$ due to the limited dynamic range of the present sample.}
The existence of a similar correlation at $z\sim 0$ \citep{bian16,
  kashino19} implies that $\la 10^8$\,$M_\odot$ galaxies in the local
universe have $n_e$ that are more similar to typical star-forming
galaxies at $z\sim 2$.

Along these lines, while \citet{sanders16b} found no correlation
between $n_e$ and $M_\ast$ for local SDSS galaxies with $M_\ast \ga
10^9$\,$M_\odot$, a limited number of studies have suggested that
$n_e$ is typically at least a factor of a few larger in metal-poor
and/or low-mass ($\la 10^8$\,$M_\odot$) galaxies compared to more
massive galaxies in the local universe (e.g., \citealt{kewley07,
  kojima20, izotov21c}; see also \citealt{kashino19}).  Such a
correlation between sSFR and $n_e$ may be expected given the
correlations between sSFR and $\Sigma_{\rm SFR}$ (e.g.,
\citealt{wuyts11, shimakawa15}), and between $\Sigma_{\rm SFR}$ and
$n_e$ (\citealt{shimakawa15, jiang19}; Figure~\ref{fig:density2}).

Consequently, the similarity in $U$ amongst local and high-redshift
galaxies at a fixed nebular abundance may be partly due to the
similarity of the sSFRs and $n_e$ (or more generally, gas density)
between these galaxies.  This conclusion is consistent with the
finding of a redshift-invariant relationship between sSFR and nebular
abundance \citep{sanders20, sanders21}: i.e., galaxies at a fixed O/H
have similar sSFRs irrespective of redshift up to $z\sim 2$.  This
finding, combined with observations that indicate a redshift-invariant
$U$ versus O/H relation \citep{topping20b, sanders20}, then imply a
redshift-invariant relationship between $U$ and sSFR.  The strong (and
apparently redshift-independent) relationship between $U$ and sSFR
(e.g., \citealt{nakajima14, sanders16b, kaasinen18}), and the fact
that sSFR positively correlates with $\Sigma_{\rm SFR}$ and gas
fraction (e.g., \citealt{reddy06a, wuyts11, genzel15, schinnerer16}),
together suggest a strong connection between $U$ and gas density
(e.g., see also \citealt{papovich22}).

Our present analysis confirms this connection.  There is ample
evidence that the increase in $U$ (and O32) with $\Sigma_{\rm SFR}$
may be driven by changes in $n_e$ (Section~\ref{sec:electrondensity};
Figure~\ref{fig:density2}).  If $n_e$ is an important factor in
modulating $U$ at $z\sim 2$, it is not unreasonable to think that
$n_e$ may play an important role in the redshift evolution of $U$ as
well.  Indeed, the redshift evolution of $U$ at a fixed stellar mass
can be most easily explained by the factor of $\simeq 10$ increases in
SFR and $n_e$ from $z\sim 0$ to $z\sim 2$.  While several previous
studies have pointed to lower gas-phase abundances at a fixed stellar
mass as being the cause of the higher $U$ inferred for high-redshift
galaxies, our results suggest that changes in gas density---which
appears to affect $n_e$ and sets the overall level of star-formation
activity---can account for much of the redshift evolution of $U$.

Aside from the redshift evolution of $U$, it is difficult to explain
the anti-correlation between $\log U$ (or O32) and $M_\ast$ at $z\sim
2$ (e.g., \citealt{sanders16b}) by metallicity effects alone.
Specifically, the stellar-mass-stellar-metallicity relation (stellar
MZR) at $z\sim 2$ \citep{strom22, kashino22} implies a decrease in
stellar metallicity of $\Delta Z_\ast \approx 0.15Z_\odot$ between
$10^{10.75}$\,$M_\odot$ and $10^{9.25}$\,$M_\odot$ galaxies at $z\sim
2$.  The BPASS model prediction shown in Figure~\ref{fig:qvszstar}
indicates that this $\Delta Z_\ast \approx 0.15Z_\odot$ translates to
a $\la 0.05$\,dex change in $\log[Q/{\rm s}^{-1}]$, which in turn
implies a $\la 0.02$\,dex change in $\log U$ assuming the scaling of
Equation~\ref{eq:U}.  This very small change in $\log U$ is clearly
insufficient to account for the $\approx 0.4$\,dex difference in the
median $\log U$ inferred between $10^{10.75}$\,$M_\odot$ and
$10^{9.25}$\,$M_\odot$ galaxies at $z\sim 2$ \citep{sanders16b}.
Hence, there must be factors other than stellar metallicity that
explain the elevated $U$ at lower stellar masses.  Indeed, previous
studies have shown that these lower-mass galaxies at $z\sim 2$ have
higher sSFRs and gas densities (e.g., \citealt{reddy06a, wuyts11,
  schinnerer16}) compared to higher-mass galaxies at the same
redshifts.

As noted above, our analysis does not conflict with previous findings
of a strong anti-correlation between $U$ and $Z_{\rm neb}$, nor does
it diminish the importance of this anti-correlation in calibrating
strong-line metallicity indicators at high redshift.  As such, these
results do not preclude the use of strong-line ratios which primarily
trace $U$ (e.g., O32, O3N2, Ne3O2) as reliable metallicity indicators
through their empirical correlation with direct measurements of
$Z_{\rm neb}$.

\section{\bf CONCLUSIONS}
\label{sec:conclusions}

We use a large sample of $z_{\rm spec} \simeq 1.9-3.7$ galaxies
selected from the MOSDEF survey to evaluate the key factors
responsible for the variation in $U$ at high redshift.  We find that
$n_e$ and $U$ correlate significantly with $\Sigma_{\rm SFR}$,
suggesting that gas density plays an important role in modulating $U$.
On the other hand, we find that $U$ is relatively insensitive to
changes in stellar metallicity or gas-phase abundance, at least
amongst galaxies in our sample.  We further find that the redshift
evolution in $U$ at a fixed stellar mass can be largely accounted for
by an increase in $n_e$ and SFR towards higher redshift.  These
results underscore the central role of gas density in explaining the
elevated $U$ inferred for high-redshift galaxies.  Measurements of
$n_e$, metallicity, and $U$ for galaxies over wider dynamic ranges in
$\Sigma_{\rm SFR}$, stellar mass, redshift, and other galaxy
properties should help to clarify the effect of gas density on the
state of the ISM throughout cosmic history.

\begin{acknowledgements}

We acknowledge support from NSF AAG grants AST1312780, 1312547,
1312764, and 1313171, grant AR13907 from the Space Telescope Science
Institute, and grant NNX16AF54G from the NASA ADAP program.  This work
made use of v2.2.1 of the Binary Population and Spectral Synthesis
(BPASS) models as described in \citet{eldridge17} and
\citet{stanway18}, and v17.02 of the Cloudy radiative transfer code
\citep{ferland17}.  We wish to extend special thanks to those of
Hawaiian ancestry on whose sacred mountain we are privileged to be
guests.  Without their generous hospitality, most of the observations
presented herein would not have been possible.

\end{acknowledgements}


\appendix

\section{Composite Spectra in Bins of $\Sigma_{\rm SFR}$}
\label{sec:appA}

\begin{figure}
  \epsscale{1.2}
  \plotone{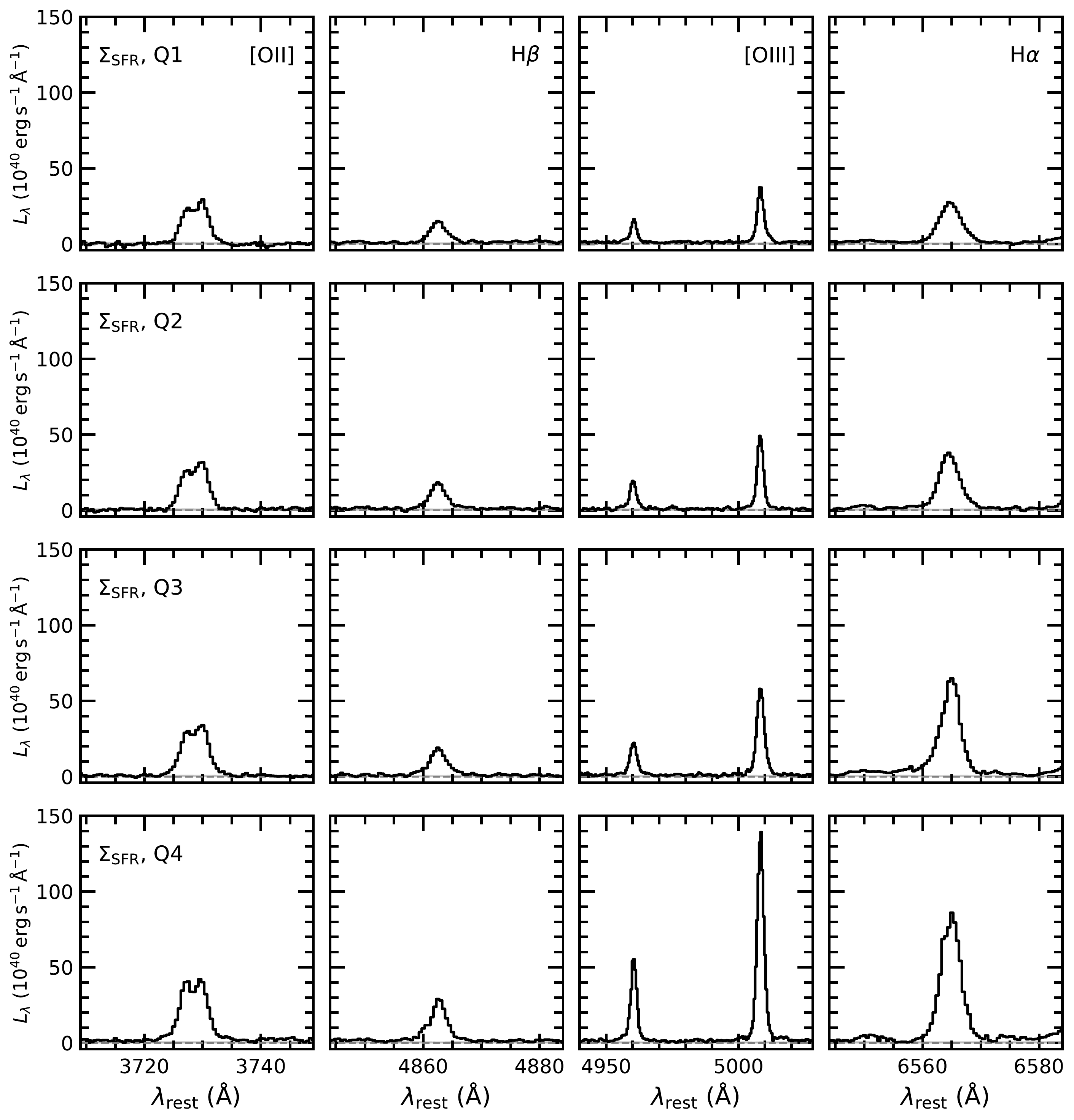}
    \caption{Composite spectra constructed from the rest-frame optical
      MOSFIRE spectra of individual galaxies in the four bins of
      $\Sigma_{\rm SFR}$, from the lowest (top row) to highest bin
      (bottom row) of $\Sigma_{\rm SFR}$.  Wavelength regions around
      $\oii$, $\hb$, $\oiii$, and $\ha$ are shown from the leftmost to
      rightmost column, respectively.}
    \label{fig:compspec}
\end{figure}

\end{document}